\documentclass[
 aip,
 amsmath,amssymb,
 reprint,
]{revtex4-1}
\usepackage{graphicx}
\usepackage{dcolumn}
\usepackage{bm}
\usepackage{xcolor}
\usepackage{soul}
\usepackage[normalem]{ulem}
\usepackage{booktabs}

\usepackage[utf8]{inputenc}
\usepackage[T1]{fontenc}
\usepackage{mathptmx}
\usepackage{etoolbox}

\makeatletter
\def\@email#1#2{%
 \endgroup
 \patchcmd{\titleblock@produce}
  {\frontmatter@RRAPformat}
  {\frontmatter@RRAPformat{\produce@RRAP{*#1\href{mailto:#2}{#2}}}\frontmatter@RRAPformat}
  {}{}
}%
\makeatother
\begin{document}

\preprint{AIP/123-QED}

\title[Designing viscoelastic fluids]{Design of model Boger fluids with systematically controlled viscoelastic properties}
\author{Jonghyun Hwang and Howard A. Stone}
\email{hastone@princeton.edu.}
\affiliation{ 
Department of Mechanical and Aerospace Engineering, Princeton University, Princeton, NJ, 08544, USA.
}%


\date{\today}

\begin{abstract}
The subject of viscoelastic flow phenomena is crucial to many areas of engineering and the physical sciences. Although much of our understanding of viscoelastic flow features stems from carefully designed experiments, preparation of model viscoelastic fluids remains a challenge; for example, fabricating a series of fluids with different fluid shear moduli $G_0$, but with an identical relaxation time $\tau$, is nontrivial. In this work, we harness the non-ideality of nearly constant-viscosity elastic fluids,  commonly known as `Boger fluids', made with polyisobutylene, to develop an experimental methodology that produces a set of fluids with desired viscoelastic properties, specifically, $G_0$, $\tau$, and the first normal stress difference coefficient $\psi_1$. Through a linear algebraic relation  between the rheological properties of interest ($G_0$, $\tau$, $\psi_1$) and the fluid compositions in terms of polymer concentration $c$, molecular weight $M_w$, and solvent viscosity $\eta_s$, we developed a `design equation' that takes $G_0$, $\tau$, $\psi_1$ as inputs and calculates values for $c$, $M_w$, $\eta_s$ as outputs. Using this method, fabrication of dilute viscoelastic fluids whose rheological properties are \textit{a priori} known can be achieved.
\end{abstract}

\maketitle

\section{\label{sec:Introduction}INTRODUCTION }

One fundamental limitation in working with polymeric materials is the difficulty in predicting or estimating their mechanical and rheological properties before the samples are fabricated. For example, varying one design parameter, such as the molecular weight of the polymer, affects both the modulus $G_0$ as well as the relaxation time $\tau$ of the  material in distinct ways.  Meanwhile, in many viscoelastic flow experiments, polymer concentration is often varied, as changing the concentration offers convenience in sample preparation and allows probing the effect of fluid elasticity. However, a change in polymer concentration not only varies $G_0$, but also affects the overall solution viscosity $\eta$, the relaxation time $\tau$, and therefore also the normal stress effects that accompany the flow of the fluid. Such common practices make it challenging to distinguish the mechanical origins of observed differences in fluid flow behavior; in other words, it may be hard to determine if an observed fluid behavior originates from differences in $G_0$, $\tau$, or the combined effects of the two, such as $G_0\tau$ or $G_0\tau^2$.

For example, many interesting viscoelastic fluid phenomena are affected or governed by the combination of $G_0$ and $\tau$. In particular,  $G_0\tau$  is typically referred to as the polymeric contribution to the solution viscosity and $G_0\tau^2$  provides a \textit{measure} of the additional normal stress generated by the flow since, for a given shear rate $\dot\gamma$, the normal stress $N_1$ generated  by the deformation of the polymer is $N_1\sim G_0\tau^2\dot\gamma^2$. Both $G_0\tau$ and $G_0\tau^2$ are combinations of dependent parameters and hence cannot be directly controlled in experiments. To gain experimental insight into the mechanical nature of polymeric fluids, material specific relationships between the independent variables and dependent variables are needed. In other words, researchers can benefit from a generalized framework that relates the rheological parameters of  polymeric fluids to the design parameters, such as polymer concentration $c$, polymer molecular weight $M$, and solvent composition and viscosity $\eta_\text{s}$. To realize this, we developed a simple method to precisely design viscoelastic fluids, in this case polymer solutions, with desired viscoelastic properties.

Characterizing mechanical and rheological properties of polymer solutions presents various challenges. For example, polymer interactions  at high concentrations of polymers make the dynamics of the fluid hard to interpret. Therefore, to highlight single polymer kinetics, studies often are in the dilute limit, where the concentration of the polymer is near-vanishing. Nonetheless, in the low polymer concentration limit, experimental measurements become challenging as a shear rheometer is often incapable of collecting reliable torque readings~\cite{ewoldt2015experimental}.

Despite these limitations, single polymer kinetics can provide important insights about the characteristics of a material, such as polymer-solvent interactions, hydrodynamic interactions, and polymer conformations~\cite{larson1989coil, schroeder2003observation}. Such knowledge can become a basis for understanding the behavior of polymers for concentrations above the dilute limit. 

Specifically, when a fluid has only a small concentration of polymer, the composition of the solution can be tuned to fabricate an idealized polymer solution whose viscosity stays nearly constant with respect to the fluid shear rate. Such materials, called `Boger fluids'~\cite{boger1977highly}, are made by mixing a dilute concentration of polymers in a compatible, high viscosity solvent. Some examples of Boger fluids that are widely used in flow experiments include (1) polyacrylamide dissolved in glycerol~\cite{jackson1984rheometrical,boger1977highly,chen2025influence}, (2) polystyrene dissolved in styrene oligomers~\cite{rothstein1999extensional,magda1993rheology,clasen2006dilute, larson2005rheology}, and (3) polyisobutylene dissolved in polybutene (high fraction of isobutylene oligomers)~\cite{verhoef1999modelling, liu2011force, su2022viscoelastic}. These fluids offer experimental and theoretical benefits by tending to decouple the effect of shear thinning from effects driven by fluid elasticity (e.g., the first normal stress difference $N_1$). Hence, Boger fluids have been used in many viscoelastic flow experiments, such as polymer flows in porous media~\cite{chen2025influence}, and in providing a simulated viscoelastic environment for active materials~\cite{liu2011force, su2022viscoelastic}. However, despite their usefulness in experiments, viscoelastic properties of Boger fluids are not \textit{a priori} known prior to the sample fabrication. For example, it remains a challenge to fabricate a series of fluids with different shear moduli $G_0$ but with an identical relaxation time $\tau$.

In this article, we use the tools available from the classical literature of fluid viscoelasticity to design Boger fluids and  systematically control their rheological properties. First, in \S\ref{sec:Material characterization}, we discuss the fabrication methods for polyisobutylene-based Boger fluids and the applicability of the Zimm model for describing the rheology. Second, in \S\ref{sec:IDENTIFYING DISCREPANCY BETWEEN THE THEORY AND A MODEL FLUID}, we identify and experimentally verify the scaling relationships~\cite{rubinstein2003polymer} between the rheological properties, i.e., shear modulus $G_0$, relaxation time $\tau$, the reciprocal of the polymer fractal dimension $\nu$, as well as the first normal stress difference coefficient $\psi_1$, and the fluid design parameters, i.e., polymer concentration $c$, polymer molecular weight $M$, and the solvent viscosity $\eta_\text{s}$. Note that $\nu$ in our work is introduced as a heuristic parameter within the Zimm framework, which is an `effective' parameter valid for the range of $M$ we explored. Lastly in  \S\ref{sec:Designing PIB Boger fluids with desired mechanical properties}, we introduce a  framework to achieve target values for $G_0$, $\tau$, and $\psi_1$ by changing the fluid composition.

The ideas presented in this work can be useful in ways other than designing Boger fluids with target viscoelastic properties. For example, our approach can provide an alternative way of inferring/estimating viscoelastic properties of fluids, whose measurements are below the torque resolution of a shear rheometer, from the measured values of the viscoelastic properties of other fluids that provide higher torque readings. Furthermore, our method can be used to test theories for viscoelastic flows that are based on Oldroyd-B or FENE-P type models, which assume weak viscoelasticity and dilute polymer concentrations. This knowledge combined with other rheometrical methods may also allow inference of other material parameters. For example, our recent work~\cite{hu2025revealing} suggests that the characterization introduced here, combined with the capillary breakup rheometry (CaBER) method, can provide a measure of extensibility of polymer molecules.

\section{MATERIAL CHARACTERIZATION\label{sec:Material characterization}}
\subsection{Material preparation and rheological measurements\label{sec:Material preparation and rheological measurements}}
We begin by introducing the ingredients of our work and our approach for characterizing  mechanical properties. The model Boger fluids are made with polyisobutylene (PIB) as the polymer, and a mixture of polybutene (PB) and light mineral oil as the solvent. Mineral oil is a convenient choice as a solvent for PIB due to its ready availability and its attractive chemical characteristics, such as being non-volatile, organic, and aromatic, which is a property that allows dissolution of the solid PIB. As an alternative to  mineral oil, kerosene oil can be used. PIB was purchased from Scientific Polymer Products Inc. at varying (viscosity-averaged) molecular weights, $M=\left[0.085, 0.4, 0.85, 1.27\right]$ $\times10^6$ (product codes $681, 040A, 682,040B$, respectively). The PB (number-averaged molecular weight; $M_\text{n}=2300$; isobutylene, >90\%; product code: 388726) and the light mineral oil were purchased from Sigma Aldrich. All materials were used as purchased. 

To verify the nominal molecular weight as well as the polydispersity, we performed gel permeation chromatography (GPC) measurements by first dissolving the polymer samples in tetrahydrofuran (Sigma Aldrich) at $1$ mg/ml, and then using standard equipment (Infinity, Agilent). In the Supplementary Material we provide the response intensity versus retention time plots for each sample from which we identified the distribution of the molecular weights. A summary of the measurements is shown in Table~\ref{tab:molecular weight}.  The PIB samples we tested are polydisperse, as indicated by the polydispersity index $PD=M_w/M_n\neq1$. However, we found that the single relaxation mode Zimm model, which considers the longest relaxation mode only, can  capture the dynamic properties of our test fluids as we discuss in \S\ref{sec:Zimm rheology} and demonstrate in Fig.~\ref{fig:compiled data}.
\begin{table*}
\caption{\label{tab:molecular weight} Values (unitness) of the various measures of molecular weight, all divided by a factor of $10^6$. $M$ in the first column denotes the nominal molecular weight value of commercially purchased polyisobutylene (PIB) products, as reported in the product description. These measurements were done with gel permeation chromatography (see Supplementary Material). Here, $M_p$ is the peak molecular weight, $M_n$ is the number-averaged molecular weight, $M_v$ is the viscosity-averaged molecular weight, $M_w$ is the weight-averaged molecular weight, and $M_z$ is the z-averaged molecular weight.}
\begin{ruledtabular}
\begin{tabular}{ccccccc}
 $M\times10^{-6}$&$M_p\times10^{-6}$&$M_n\times10^{-6}$&$M_v\times10^{-6}$ & $M_w\times10^{-6}$& $M_z\times10^{-6}$&$PD=M_w/M_n$\\
\hline
0.085&  0.096&  0.0612& 0.095& 0.101& 0.153&1.656
\\
0.4&  0.570& 0.257&  0.457& 0.491& 0.732&1.914
\\
0.85&  0.921& 0.485&  0.744& 0.786& 1.069&1.621
\\
1.27\footnote{Nominal value for this sample lies outside the calibration range of GPC. Hence, measurements made with this material were not considered when identifying the molecular weight dependency in \S\ref{sec:Molecular weight dependence}.}&  1.027& 0.383&  0.772& 0.830& 1.201&2.167\\
\end{tabular}
\end{ruledtabular}
\end{table*}

To fabricate the test fluids, PIB was first dissolved in light mineral oil, using a magnetic stirrer, to make $3-5$ wt\% stock solutions. The dissolution, at an elevated temperature of  80 $^{\circ}$C, took $3-10$ days, depending on the molecular weight of the solute (PIB). PIB can degrade under extreme thermal stimuli~\cite{srinivasan1985thermal}, hence we chose to keep the temperature below 100 $^\circ$C. Similarly, PB and mineral oil were mixed at  4:1 weight ratio to prepare a reduced-viscosity stock because 100\% PB is hard to handle due to its high viscosity. To make the final sample, the PIB-mineral oil stock solution and the PB-mineral oil mixture were mixed at desired ratios. To ensure reproducibility, the typical variance in the measured weights of each ingredient were maintained within 0.1\% of the targeted value. The produced samples maintained their viscoelastic properties for at least three months of storage in sealed containers at room temperature.

We performed small-amplitude-oscillatory shear (SAOS) measurements and shear viscosity measurements with an MCR 302e rheometer (Anton Paar, USA) using the cone-plate geometry. The cone had an angle of $1^{\circ}$, diameter of 50 mm, and a truncation length of 50 $\mu$m at the tip. The plate was maintained at 22 $^{\circ}$C during the measurements. The frequency sweep measurements probed the fluid responses from $f=\omega/2\pi=10^{-2}$ Hz to $10^2$ Hz at $30-100$ \% shear amplitude (here, $\omega$ is the angular frequency of the oscillation). During the SAOS procedure, we ensured that measurements were done in the linear viscoelastic regime; the solutions remained in the linear viscoelastic regime up to nearly $100$ \% shear amplitude with a few exceptions at either high $M$ or $c$. The rotational shear measurements were taken at varying shear rates $\dot\gamma = 10^{-1} - 3\times10^3$ s$^{-1}$. For each condition, rheological properties of two or three independently prepared solutions were measured.

\subsection{\label{sec:Characterizing the dilute to semi-dilute transition}Characterizing the dilute to semi-dilute transition}
\begin{figure*}
\includegraphics[width=170mm]{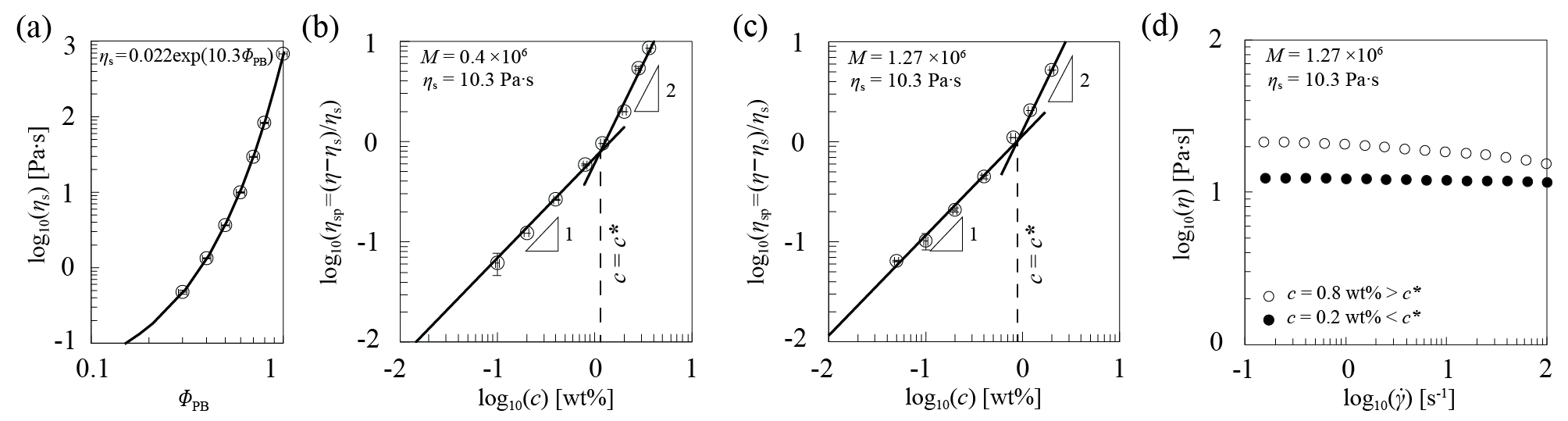}
\caption{\label{fig:lineariy plot}Rheology of dilute Boger-like viscoelastic fluids. (a) Solvent viscosity $\eta_\text{s}$ as a function of PB weight fraction, $\phi_{\text{PB}}$, in a PB-light mineral oil mixture. We  found empirically that $\eta_\text{s}\approx 0.022\exp(10.3\phi_{\text{PB}})~\left[\text{Pa}\cdot\text{s}\right]$. (b) and (c) For a fixed $\phi_{\text{PB}}$, we measured the specific viscosity $\eta_{\text{sp}}=(\eta-\eta_\text{s})/\eta_\text{s}$ as a function of PIB concentration $c$ (wt\%), where $\eta$ is the measured solution viscosity in the low shear rate limit ($\dot\gamma=0.1$ s$^{-1}$). For $c<c^*$, where $c^*$ denotes the critical overlap concentration where neighboring polymers start to interact, $\eta_{\text{sp}}\propto c$, as expected for a dilute viscoelastic fluid. $c^*$ depends on molecular weight $M$ of the polymer;  in the range of different polymer molecular weights we investigated in this study, $c<0.6$ wt\% could generally be considered dilute. The indicated line with slope 2 is the theoretical prediction of $\eta_{\text{sp}}$ versus $c$ in the semi-dilute unentangled regime for a polymer solution in the $\theta$-solvent condition ($\nu=1/2$). (d) Two shear viscosity measurements of $M= 1.27\times10^6$ PIB at two different concentrations, $c=0.2<c^*$ and $c=0.8\approx c^*$. When $c\approx c^*$, polymer shear thinning starts to become non-negligible. Horizontal error bars are standard 5\% deviations and the vertical error bars were measured.}
\end{figure*}

To systematically control the solvent viscosity, we mixed PB and the light mineral oil at varying weight ratios, as summarized in Fig.~\ref{fig:lineariy plot}(a) and Table~\ref{tab:solvent viscosity}. We observed that the solvent viscosity, $\eta_\text{s}$, follows an exponential dependence on $\phi_{\text{PB}}$, the weight fraction of PB in the PB and light mineral oil mixture,
\begin{equation}
  \eta_\text{s}\approx0.022 \exp(10.3\phi_{\text{PB}}) ~\left[\text{Pa}\cdot\text{s}\right].  
\end{equation}
Using this empirical formula, solutions with a desired solvent viscosity could be prepared.
\begin{table}
\caption{\label{tab:solvent viscosity} Solvent viscosity with varying PB weight fraction $\phi_{\text{PB}}$.}
\begin{ruledtabular}
\begin{tabular}{cccc}
 Light mineral oil [wt\%] &PB [wt\%]&$\eta_\text{s}$ [Pa$\cdot$s]&range\footnote{Half of the difference between the maximum and minimum of the measured values of $\eta_\text{s}$} [Pa$\cdot$s]\\
\hline
0  &  100&  680&n/a\footnote{Measured once  to avoid excessive torque on the machine}\\
20&  80  & 84.6& $\pm$1.05\\
30&  70  & 28.9& $\pm$0.48\\
40&  60& 10.3& $\pm$0.11\\
50&  50& 3.62& $\pm$0.01\\
60&  40& 1.36& $\pm$0.01\\
70&  30& 0.51& $\pm$0.03\\
\end{tabular}
\end{ruledtabular}
\end{table}

Next, to understand the effect of the polymer concentration on the solution viscosity, we added small amounts of PIB to the solution and performed rotational shear tests. In these measurements, the weight ratios between PB and mineral oil were kept at a chosen value, while the weight percent of PIB was measured relative to the total solution weight. Let $\eta$ denote the total solution viscosity in the low shear rate limit ($\dot\gamma=0.1$ s$^{-1}$).
The difference  $\eta-\eta_\text{s}\equiv\eta_\text{p}$ accounts for the polymer contribution to the total solution viscosity. In Fig.~\ref{fig:lineariy plot}(b, c), we report measurements of the specific viscosity $\eta_{\text{sp}}=(\eta-\eta_\text{s})/\eta_\text{s}$ versus the PIB concentration $c$, in wt \%, for two different PIB molecular weights, with panel (b) $M=0.4\times10^6$ and panel (c) $1.27\times10^6$.

We observe that at sufficiently low polymer concentrations, typically $c<0.8$ wt \% for $M=1.27\times10^6$, $\eta_{\text{sp}}$ increases linearly proportional to $c$. This suggests that at low polymer concentrations, changes to the solution viscosity due to the presence of the polymer can be expressed as a first-order expansion in the polymer concentration~\cite{liu2009concentration}, i.e., $\eta/\eta_\text{s}\approx 1+Ac+\cdots$, where $A$ is a proportionality constant. In this state, the polymer solution is considered `dilute' and the polymer molecules are assumed to not interact with each other. However, as $c$ approaches the critical overlap concentration $c^*$, which signifies the transition from the dilute regime to the semi-dilute unentangled regime, $\eta_{\text{sp}}$ starts to increase nonlinearly with increasing polymer concentration~\cite{del2017relaxation}. In this regime, the specific viscosity is expected to scale with concentration as $\eta_\text{sp}=\left(c/c^*\right)^{1/(3\nu-1)}$, which gives $\eta_\text{sp}\propto c^2$ for $\nu=1/2$. Although we were unable to prepare solutions with $c>2$ wt\% for $M=1.27\times10^6$ and $c>4$ wt\% for  $M=1.27\times10^6$, a few  data points above $c=c^*$ suggest that our solutions show an effective $\nu$ that is close to $1/2$. We provide an in-depth discussion regarding $\nu$ in \S\ref{sec:Zimm rheology}. For $c>c^*$, shear-thinning effects are non-negligible, and the polymer solution is therefore no longer considered a Boger fluid, as shown in Fig.~\ref{fig:lineariy plot}(d). Note that $c^*$ can vary depending on the polymer molecular weight as well as the solvent viscosity. For the measurements shown in Fig.~\ref{fig:lineariy plot}(d), we used $\eta_s\approx 10.3$ Pa$\cdot$s. All viscoelastic fluids presented and discussed in the following sections are considered dilute.

\subsection{\label{sec:Zimm rheology}Zimm rheology}
Many of the widely used formulations of Boger fluids exhibit viscoelastic characteristics that follow predictions of the Zimm model~\cite{zimm1956dynamics}, which assumes a solution with a dilute polymer concentration, where each polymer chain is composed of hydrodynamically interacting beads connected by springs.
Therefore, the change in the solution viscosity due to the presence of this `polymer ball' is sensitive to the radius of gyration of the polymers, and this size effect is captured by the inverse of the polymer fractal dimension, $\nu$. We note that the Rouse model, which is often discussed in conjunction with the Zimm model, is not applicable in our study, as the Rouse model is more appropriate to discussions of polymer melts. 

There are many polymer solutions that are known to exhibit Zimm-like dynamics. For example, experiments show that dynamic moduli, $G^*(\omega)=G'(\omega)+iG''(\omega)$ where  $\omega$ is the angular shear frequency, of polyisobutylene-polybutene-based Boger fluids closely follow the three-parameter model that Zimm theory describes, comprised of a shear modulus $G_0$, Zimm relaxation time $\tau$, and the reciprocal of the polymer fractal dimension $\nu$ introduced in the previous paragraph~\cite{more2023rod}. 
However, fitting a three-parameter model to a set of two curves, e.g., storage $G'(\omega)$ and loss $G''(\omega)$ moduli, leaves room for ambiguity; multiple sets of distinct values for $G_0, \tau$, and $\nu$ can yield acceptable fitting results. One way to avoid this multiplicity is to select an appropriate combination of the polymer and solvent with an \textit{a priori} known value of $\nu$. For example, polystyrene dissolved in dioctyl phthalate is at the $\theta$-condition~\cite{johnson1970infinite} at 22 $^\circ$C, which constrains $\nu=1/2$. However, this method relies on the availability of such known chemical compositions, and often $\theta$-solvents can be toxic, volatile, or hard to work with.

\begin{figure*}
\includegraphics[width=170mm]{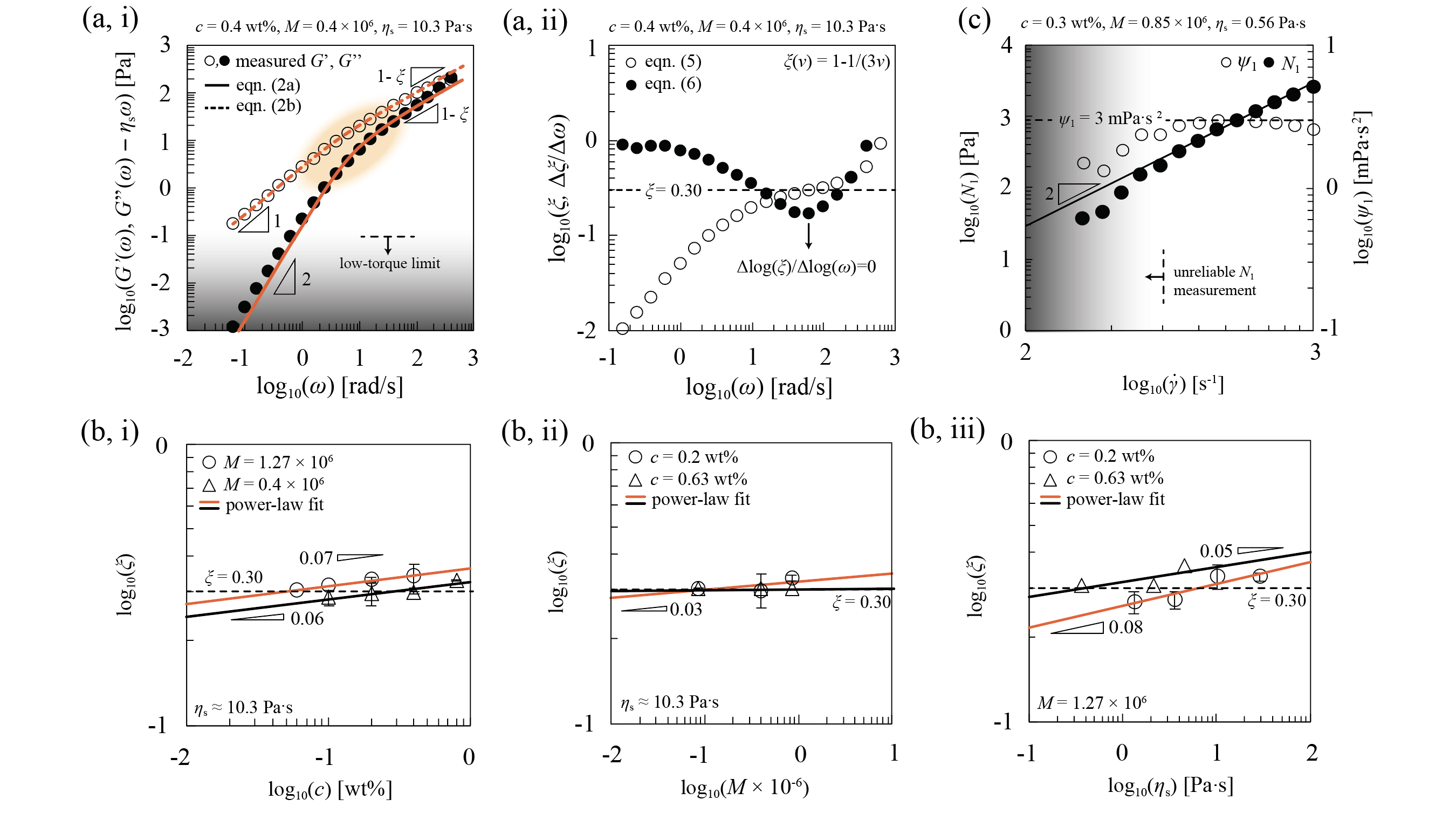}
\caption{\label{fig:fitting method} Zimm model curve fitting method to identify $G_0, \tau$, and $\xi$. 
a, \textit{i}) For demonstration purposes, a measurement of the storage ($G'$) and loss moduli ($G''-\eta_\text{s}\omega$) of $0.4$ wt\% of $0.4\times10^6$ PIB dissolved in $40$ \% mineral oil mixed with $60$ \% PB solution is presented. 
a, \textit{ii}) Using Eq.~(\ref{eqn:fitting chi}), $\xi$ can be found by looking at the region where $\Delta\log(\xi(\omega))/\Delta\log(\omega)$ is the smallest. In this example presented, $\eta_\text{s}=10.3~\text{Pa}\cdot\text{s}, G_0 =2.2 ~\text{Pa}, \tau=1.05 ~\text{s},$ and $\xi=0.30$. 
b) Fitted $\xi$ in relation to the design parameters, $c, M$, and $\eta_s$. $\xi$ shows weak dependence on $c$ and $\eta_s$, understandably due to changes in the fluid composition, but no appreciable dependence on $M$; the mean value  is $\xi_{\text{avg}}\approx0.30$. Different symbols and colors represent different $M$ and $\eta_\text{s}$. 
c) From the rotational shear measurements, we determined $\psi_1$ by identifying the region where the first normal stress difference $N_1$ increases as $N_1\propto\dot\gamma^2$. }
\end{figure*}

Specifically, the Zimm model expresses the storage and loss moduli $G'(\omega)$ and $G''(\omega)$ of a fluid as~\cite{rubinstein2003polymer}
\begin{subequations}
\label{zimm dynamic modulus}
\begin{equation}
G'(\omega)=G_0\frac{ \omega\tau~\text{sin}\left((1-1/(3\nu)) ~\text{arctan}(\omega\tau)\right)}{[1+(\omega\tau)^2]^{(1-1/(3\nu))}/2},
\end{equation}
\begin{equation}
G''(\omega)-\eta_\text{s}\omega=G_0 \frac{\omega\tau~\text{cos}\left((1-1/(3\nu)) ~\text{arctan}(\omega\tau)\right)}{[1+(\omega\tau)^2]^{(1-1/(3\nu))}/2}.
\end{equation}
\end{subequations}
To enhance readability, going forward we will denote $\xi=(1-1/(3\nu))$. Eq.~(\ref{zimm dynamic modulus}) predicts that at high oscillation frequencies ($\omega \tau\gg 1$) the moduli vary as 
\begin{subequations}
\label{zimm high frequency}
\begin{equation}
\lim_{\omega\tau \gg1} G'(\omega)\approx G_0~\text{sin}\left(\frac{\pi}{2}\xi\right)(\omega\tau)^{1-\xi},
\end{equation}
\begin{equation}
\lim_{\omega\tau \gg1} G''(\omega)-\eta_\text{s}\omega \approx G_0~\text{cos}\left(\frac{\pi}{2}\xi\right)(\omega\tau)^{1-\xi}.
\end{equation}
\end{subequations}
Likewise, at low frequencies ($\omega\tau \ll 1$), the asymptotic behaviors of $G'$ and $G''$ are
\begin{subequations}
\label{eqn:zimm low frequency}
\begin{equation}
\lim_{\omega\tau \ll 1} G'(\omega)\approx G_0\xi(\omega\tau)^{2},
\end{equation}
\begin{equation}
\lim_{\omega\tau \ll 1} G''(\omega)-\eta_\text{s}\omega \approx G_0(\omega\tau).
\end{equation}
\end{subequations}
These asymptotic limits suggest that with an appropriate method for fitting the experimentally measured curves of $G'$ and $G''$ as a function of $\omega$, the three control parameters ($G_0, \tau, \nu ~\hbox{or}~ \xi$) can be identified independently. To demonstrate this approach to material property measurement, we prepared a solution of $0.2$ wt\% PIB dissolved in a $40$ \% light mineral oil and $60$ \% PB ($\eta_\text{s}\approx 10.3$ Pa$\cdot$s), whose storage and loss moduli were measured and presented in Fig.~\ref{fig:fitting method}(a). 

The process we used to identify the shear modulus $G_0$, the longest relaxation time $\tau$ (Zimm relaxation time), and the $\nu$-dependent parameter in the Zimm model $\xi(\nu)=1-1/(3\nu)$ is as follows. We first begin by identifying $\xi$. For both of the $G'$ and $G''$ curves, Eq.~(\ref{zimm high frequency}) predicts $G', G''-\eta_s\omega\propto \omega^{1-\xi}$ for $\omega\tau\gg1$. However, the fitting of  $G''-\eta_s\omega$ is sensitive to the quality of the measurement of the solvent viscosity, and the $G'$ data tend to curve upward at high values of $\omega$ instead of increasing at a constant rate (see Fig.~\ref{fig:fitting method}(a,i)). We point to multiple possibilities concerning why $G'$ and $G''$ curves cross-over at a high frequency. First, inertial effects can influence measurements at high shear frequencies~\cite{ewoldt2015experimental}. On the other hand, one study~\cite{clasen2006dilute} suggests that such a variation indicates the existence of shorter relaxation modes coming from the weak elasticity of the oligomeric solvent. Lastly, at sufficiently high frequencies $G'$ and $G''$ curves are expected to crossover due to the finite chain-length effects (see Fig.~\ref{fig:fitting method}(a,i) at $\omega\approx 600$ rad/s), at frequencies higher than we can access with our rheometer. With shorter chains, such crossover occurs at lower frequencies, which delimits the range over which $G'$ and $G''$ increases with $\omega^{1-\xi}$. These features make power-law fitting challenging in solutions made with low polymer concentrations or low polymer molecular weights, because the window of the data points with a consistent power-law increase becomes narrower. 

Instead, we examine the ratio of $G'$ and $G''$ at high frequencies to find
\begin{equation}\label{eqn:fitting chi}
  \xi=\lim_{\omega\tau \gg 1}\xi(\omega)=\lim_{\omega\tau \gg 1}\frac{2}{\pi}\tan^{-1}\left(\frac{G'}{G''-\eta_\text{s}\omega}\right),
\end{equation}
and identify a region where the ratio $\xi(\omega)=\frac{2}{\pi}\tan^{-1}\left(\frac{G'}{G''-\eta_\text{s}\omega}\right)$ remains at a constant value. An example of applying this method is illustrated in Fig.~\ref{fig:fitting method}(a,~\textit{ii}). Since $\tan( \xi(\omega))$ is expected to plateau at a constant value when both $G'$ and $G''$ increase at the same rate, we seek the point where $d \xi(\omega)/d\omega=0$, the local minima of the curve. 

Since the measurements were done with values of $\omega$ that are equally spaced in log space, we applied finite differencing to the log-valued $ \xi(\omega)$ and $\omega$ to find the value of $ \xi(\omega)$ where the slope is a minimum, i.e., 
\begin{equation}
\min\left[\frac{\Delta\log(\xi(\omega))}{\Delta\log(\omega)}=\frac{\log(\xi(\omega_{i+1})/\xi(\omega_{i-1}))}{\log(\omega_{i+1}/\omega_{i-1})}\right],
\end{equation}
for $i=1,...,n$ indicating the number of data points measured. 

In Fig.~\ref{fig:fitting method}(b,~\textit{i}-\textit{iii}). we show the fitting results of $\xi$ as function of $c$, $M$, and $\eta_s$. $\xi$ exhibits a weak power-law dependence on both $c$ and $\eta_s$, but not on $M$, which is understandable as only $c$ and $\eta_s$ change the overall fluid composition. The values of $\xi$ stayed around the mean value $\xi_{\text{avg}}\approx0.30$ with a standard deviation of $\pm0.023$, with a couple of outliers. This small deviation suggests that we can treat $\xi$ as a constant in our measurement scheme. Hence, $\xi(\nu)=1-1/(3\nu) \approx0.3$ corresponds to $\nu= 0.47$.

What does this value of $\nu=0.47$ tell us about the fluid? For an ideal polymer chain dissolved in a theta solvent, $\nu=1/2$. If instead a polymer chain is dissolved in a chemically identical solvent (e.g., high molecular weight polyisobutylene dissolved in isobutylene oligomer solvents), the system would be considered athermal~\cite{macdonald1997shear}, and one would expect~\cite{rubinstein2003polymer} $\nu=3/5$. However, note that we use $\nu$ as a standard Zimm-model heuristic, and, therefore, we regard that $\nu$ in our work is an `effective' parameter that is only valid for the range of $M$ we explored at a constant temperature. Most experimentally accessible polymer solutions reside in the thermal crossover regime between the $\theta$- and athermal solvent limits (intermediate solvent quality, which includes our PIB-PB-mineral oil system) their behavior is more generally described by the solvent-quality parameter $z=k_0(1-\theta/T)\sqrt{M_w}$, which is a function of dimensionless temperature $\theta/T$, molecular weight $M_w$, and a chemistry dependent constant $k_0$, rather than by a single universal exponent $\nu$~\cite{schroeder2018single,prakash2019universal}. 

The value of $\nu=0.47$ obtained with our sample material at $22~^\circ$C indicates that our system is not an ideal athermal solvent. This implies that the addition of mineral oil drives the solvent toward being a poor solvent for the polymer. This observation is best visualized by the data shown in Fig.~\ref{fig:fitting method}(b, iii), where $\xi=1-1/(3\nu)$ increases with $\eta_s$, meaning that $\nu$ moves away from the athermal limit and is driven `poor' as more mineral oil is added. A quantitative determination of how $\nu$ changes with the solvent composition for this ternary mixture (PIB-PB-mineral oil) would require independent coil-dimension measurements over a wider range of temperatures and molecular weights. Such an experimental program lies beyond the scope of our work, which focuses on developing a rheology-based practical fluid design scheme, yet we suggest that this would be a natural direction for future work (we acknowledge an anonymous referee for this feedback). 

Next, the values of $G_0$ and $\tau$ can be fitted from the data  in the intermediate to low frequency regime. With $\xi$ determined, the free parameters are reduced to two ($G_0,\tau$), and so next we fit Eqs.~(\ref{zimm dynamic modulus}) to the curves of $G'$ and $G''$. Because the $G'$ measurements in the low-$\omega$ regime were unreliable due to the low-torque limit of our rheometer, we instead chose the values of $G_0$ and $\tau$ that best fit the data points in the intermediate-$\omega$ regime, where we observe the transition from $G'\propto \omega^2$ to $G'\propto\omega^{1-\xi}$ (region highlighted in orange in Fig.~\ref{fig:fitting method}(a)). 

Lastly, although not frequently discussed to our knowledge in most of the literature on Zimm-like fluids, the first normal stress difference $N_1$ can be related to the oscillatory measurements by an approximation in the low shear frequency limit~\cite{bird1987dynamics, more2023rod},
\begin{equation}\label{eqn:Normal stress approximation}
\psi_{1, ~\text{ideal}}\equiv\lim_{\dot\gamma\rightarrow0}\frac{N_1}{\dot\gamma^2}\approx \lim_{\omega\rightarrow0}\frac{2G'}{\omega^2}\approx 2\xi G_0\tau^2.
\end{equation} 
The above expression is obtained by the direct substitution of Eq.~(\ref{eqn:zimm low frequency}a) (valid for $\omega\tau\ll1$) into $\lim_{\omega\rightarrow0}G'$. We distinguish this theoretical estimation of the first normal stress difference coefficient with a subscript `ideal'. This approximation, however, does not best capture the measured values of the first normal stress difference coefficient $\psi_1$. Hence, we include another fitting constant $\delta$ in Eq.~(\ref{eqn:Normal stress approximation}), such that
\begin{equation}
    N_1\approx2\delta \xi G_0\tau^2\dot\gamma^2=\delta\xi\psi_{\text{OB}} \dot\gamma^2,
\end{equation}
where $\psi_{\text{OB}}=2G_0\tau^2$ is the first normal stress difference coefficient predicted with the Oldroyd-B model. Hence, the factor $\delta\xi$ signifies the deviation of the real value from an ideal limit. Often, measurements of $\psi_1$ plotted on a log-log scale are  compared directly with Eq.~(\ref{eqn:Normal stress approximation}), which provides a seemingly acceptable fit~\cite{more2023rod}. However, we found that $\delta$ was typically on the order of $\xi$ and in the range $0.2-0.4$, and hence when $\psi_1$ is plotted on a linear scale, it is lower than what Eq.~(\ref{eqn:Normal stress approximation}) predicts. We are not the first to notice this deviation, as evidenced by numerous literature that address the difference between the measured $\psi_1$ and that predicted from the linear viscoelastic parameters~\cite{james2021pressure, sharma2012intriguing,rothstein1999extensional}. Here we note that $\delta$ may originate from the limitations in applying the linear-viscoelastic limit of the Laun's rule, non-affine motions (e.g. rolling and tumbling) of polymers~\cite{schroeder2005characteristic}, or alterations to local hydrodynamic interactions due to a non-ideal solvent. Going forward, we will denote the first normal stress difference coefficient as
\begin{equation}\label{eqn:our def of psi1}
    \psi_1\equiv 2\delta\xi G_0\tau^2=\delta\xi\psi_{\text{OB}},
\end{equation}
and will discuss how $\delta\xi$ varies with the design parameters of the fluid in \S\ref{sec:IDENTIFYING DISCREPANCY BETWEEN THE THEORY AND A MODEL FLUID}. In measurements, we find $\psi_1$ by identifying the region where $\psi_1$ is nearly constant with respect to changes in $\dot\gamma$, which corresponds to the region where $N_1\propto \dot\gamma^2$, as shown in Fig.~\ref{fig:fitting method}(c). 

\begin{figure*}
\includegraphics[width=170mm]{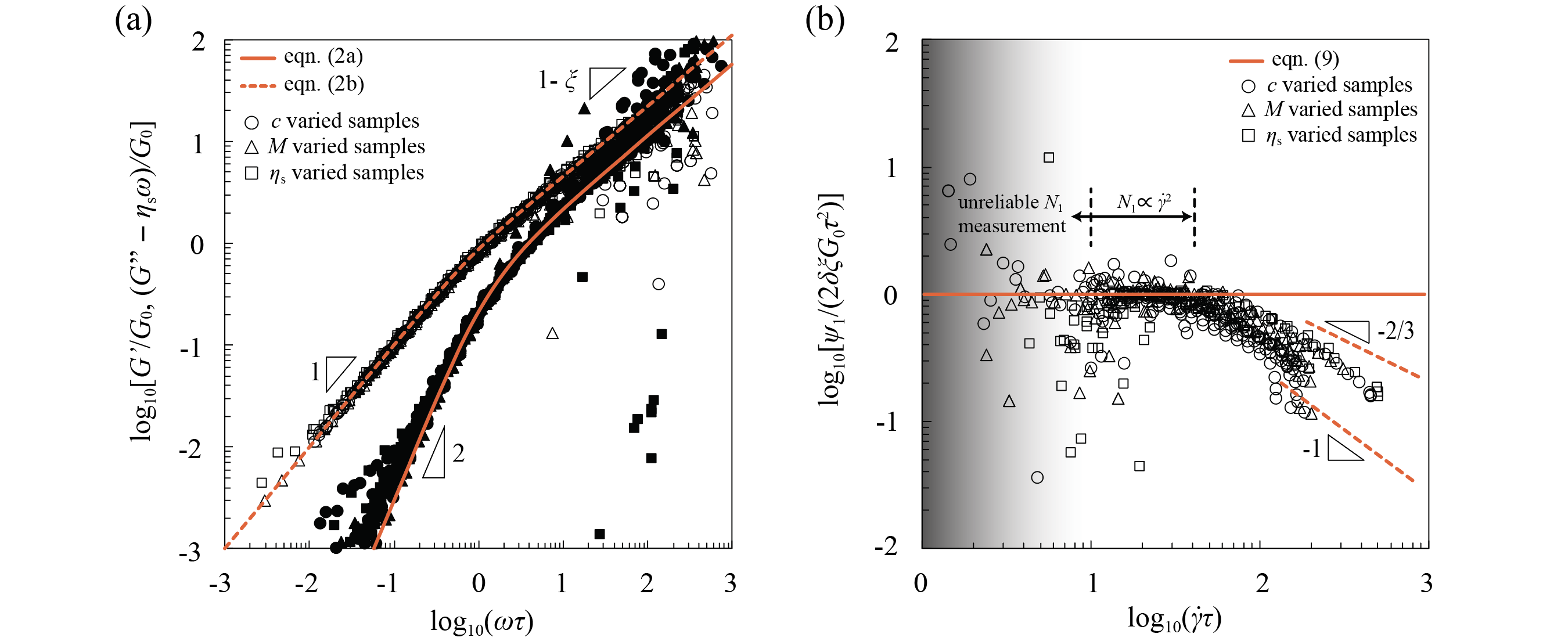}
\caption{\label{fig:compiled data} Measured rheological data normalized by the fitted parameters. a) Compiled data of all sets of SAOS measurements with fitted $G_0, \tau$ and $\xi$. All data points collapse onto the Zimm model normalized by $G_0$ in the vertical axis and by $1/\tau$ in the horizontal axis. Despite the polydispersity of the sample polymers used, considering only the longest relaxation mode can well capture the trends in dynamic moduli. b) Measurement of $\psi_1$ normalized by $2\delta\xi G_0\tau^2$ versus $\dot\gamma\tau$. Values of $G_0, \tau$, and $\xi$ are obtained from SAOS measurements, and $\psi_1$ and $\delta$ are separately obtained from the rotational shear tests (see (d) panels of Figs.~\ref{fig:c-dependence}, \ref{fig:M-dependence}, and \ref{fig:eta-dependence}). Typically when $10<\dot\gamma\tau<30$, $N_1\propto \dot\gamma^2$ ($\psi_1$ is nearly constant). Although the measurements below $\dot\gamma\tau=10$ are unreliable due to the limitations in the normal force reading resolution of the rheometer, it is expected that $N_1\propto\dot\gamma^2$ in that regime too. However, when $\dot\gamma\tau>30$, $\psi_1$ starts to decrease,  following a power-law with an exponent $\approx -2/3$.}
\end{figure*}

To demonstrate consistency of our fitting scheme in identifying the rheological parameters $G_0,\tau,{\nu},\psi_1,$ and $\delta$, we compiled all of our measurements in a master plot shown in Fig.~\ref{fig:compiled data}(a). The results show noise in measurements at both low $\omega$ and high $\omega$, whose origins have been discussed above. Thus, we claim that the three-parameter Zimm model is sufficient for capturing the key viscoelastic parameters, $G_0, \tau,$ and $\nu$, that govern the mechanical properties of our test fluids. We also show the data obtained from the rotational shear measurements in Fig.~\ref{fig:compiled data}(b), where we identified the parameters $\psi_1$ and $\delta$.

\section{Characterizing the theory-experiment discrepancies}\label{sec:IDENTIFYING DISCREPANCY BETWEEN THE THEORY AND A MODEL FLUID}
In \S \ref{sec:Material characterization}, we described a method for producing a dilute viscoelastic solution using three readily available chemical components, PIB, PB, and light mineral oil. Also, we discussed how the three parameters in the Zimm model, $G_0, \tau$, and $\xi$, can be identified consistently based on experimental measurements of the viscoelastic moduli. In this section, we represent the dependence of the rheological parameters ($G_0, \tau, \psi_1$) on the fluid compositions ($c, M, \eta_s$) using power-law forms of the type $Y\propto X^\alpha$, where $\alpha$ are expected to be related to the solvent quality, i.e., $\nu$   (see Table~\ref{tab:summary table}). 

Furthermore, we will discuss the scaling concepts and theoretical predictions for the dynamics of dilute polymer solutions and how those predictions deviate from the behaviors observed in the model material. We find that while the scaling theory generally captures the trend, appreciable differences remain between the theoretical predictions and the behavior of the model material. We refer to this departure from the ideal model predictions as `non-ideality'. The non-ideality of the model experimental system becomes the basis of our fluid design scheme, and we show how we harness this non-ideality to control  independently the rheological parameters of interest.

We emphasize that the empirically found exponents that appear in this section are effective exponents obtained over a finite range of the polymer molecular weights and at a fixed temperature specifically for the PIB in PB–mineral oil system, and hence should not be taken as universal exponents. Our use of the scaling relations is intended as a practical framework to construct a design matrix (Eq.~\ref{eqn: matrix form exp inverted}). In other words, we use the Zimm model (Eq.~\ref{zimm dynamic modulus}) as a functional form to parametrically represent the dynamic moduli of our material. The conclusions of our work do not rely on identifying any universality of the fitted parameters.

\subsection{Dependence on polymer concentration \label{sec:Concentration dependence}}
\begin{figure}
\includegraphics[width=85mm]{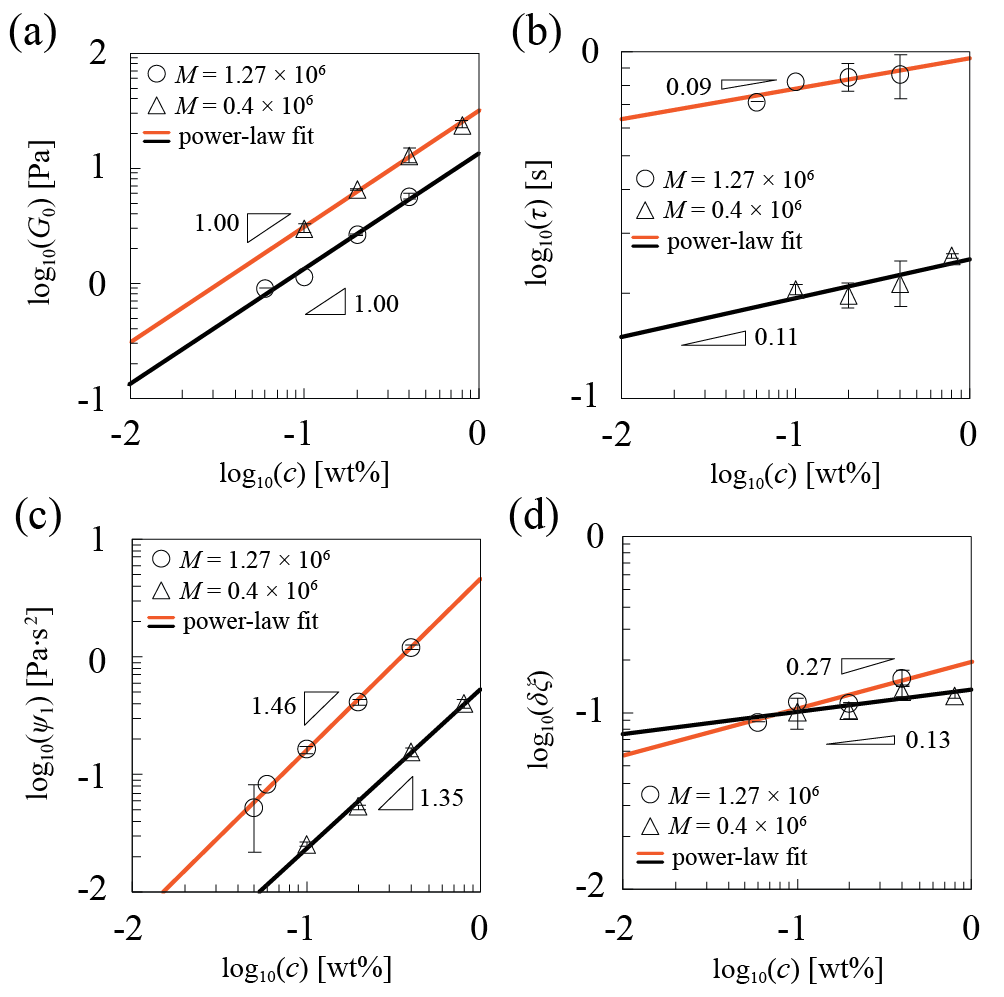}
\caption{\label{fig:c-dependence} Measured power-law dependence of a) $G_0$, b) $\tau$, c) $\psi_1$, and d) $\delta\xi$ on $c$, for constant $M=[0.4,1.27]\times 10^6$  and constant $\eta_s\approx 10$ Pa$\cdot$s. Refer to Tables~\ref{tab:summary table} and \ref{tab:summary table exp} for theory/experiment comparison. All data points shown in this figure were measured with solutions whose polymer concentrations were considered dilute.
}
\end{figure}
To identify the polymer concentration dependence of $G_0, \tau$,  $\psi_1$, and $\delta\xi$, we fixed the solvent viscosity $\eta_\text{s}$ and the polymer molecular weight $M$, and varied only $c$. The fitted values of  $G_0, \tau, \psi_1,\delta\xi=\psi_1/\psi_{\text{OB}}$ as functions of $c$ are shown in Fig.~\ref{fig:c-dependence}(a-d). Here, $\psi_{\text{OB}}=2G_0\tau^2$ is not independently measured but is comprised of the fitted values $G_0$ and $\tau$. The scaling theory predicts that both $G_0$ and $\psi_1$ scale linearly with $c$, while $\tau$ is independent of $c$. Also, the Zimm theory considers a single coil in an infinite solvent, and hence $\xi$ is assumed constant.

Specifically, $G_0$ in the Zimm theory is expressed as a function of the Kuhn length $b$, degree of polymerization $N$, polymer volume fraction $\phi$, and thermal energy scale $k_BT$, i.e., $G_0=k_BT\phi/(Nb^3)$. Since $\phi$ is proportional to $c$ in a constant-density fluid, we expect that $G_0\propto c$, and this prediction is confirmed by the oscillatory shear measurements presented in Fig.~\ref{fig:c-dependence}(a). 

The Zimm relaxation time $\tau$ is (theoretically) expected to remain constant for dilute polymer concentrations where polymers are assumed to not interact with each other~\cite{del2017relaxation, rubinstein2003polymer}. However, numerous papers report~\cite{del2017relaxation} that even when $\eta_\text{sp}\propto c$ (dilute regime), $\tau$ increases weakly with $c$. Our measurements likewise indicate a weak dependence, $\tau\propto c^{0.10\pm0.01}$, as shown in Fig.~\ref{fig:c-dependence}(b). The exponent of 0.1 will be system dependent and can stem from the  differences in the polymer molecular weight, stiffness, and solvent types; e.g., the exponent is approximately $0.76$, for a dilute polystyrene-dioctyl phthalate system~\cite{del2017relaxation}. Similarly, experiments show $\xi$ varies with $c$. Although we do not comment on why both $\xi$ and $\tau$ increase with $c$, this may suggest that polymer-polymer interactions are playing some role.

The rotational shear measurements show that $\psi_1\propto c^{1.41\pm0.06}$. Considering that $\tau\propto c^{0.10\pm0.01}$ and that $\psi_1\propto G_0\tau^2$, the semi-empirical prediction gives $\psi_1\propto c^{1.2}$. This difference suggests that the factor of $\delta\xi=O(0.1)$ is $c$-dependent, as shown in Fig.~\ref{fig:c-dependence}(d). In the plot, we present $\delta\xi$, not just $\delta$, to assess their combined effect on $\psi_1$. Therefore, we show that Eq.~(\ref{eqn:Normal stress approximation}) would over-predict $N_1$ by a factor of $\delta^{-1}$. Interestingly, $\psi_\text{OB}=\psi_1/\delta\xi\propto c^{1.22}$, which is close to the semi-empirical prediction of $\psi_1$ on $c$.

\subsection{Dependence on polymer molecular weight \label{sec:Molecular weight dependence}}
\begin{figure}
\includegraphics[width=85mm]{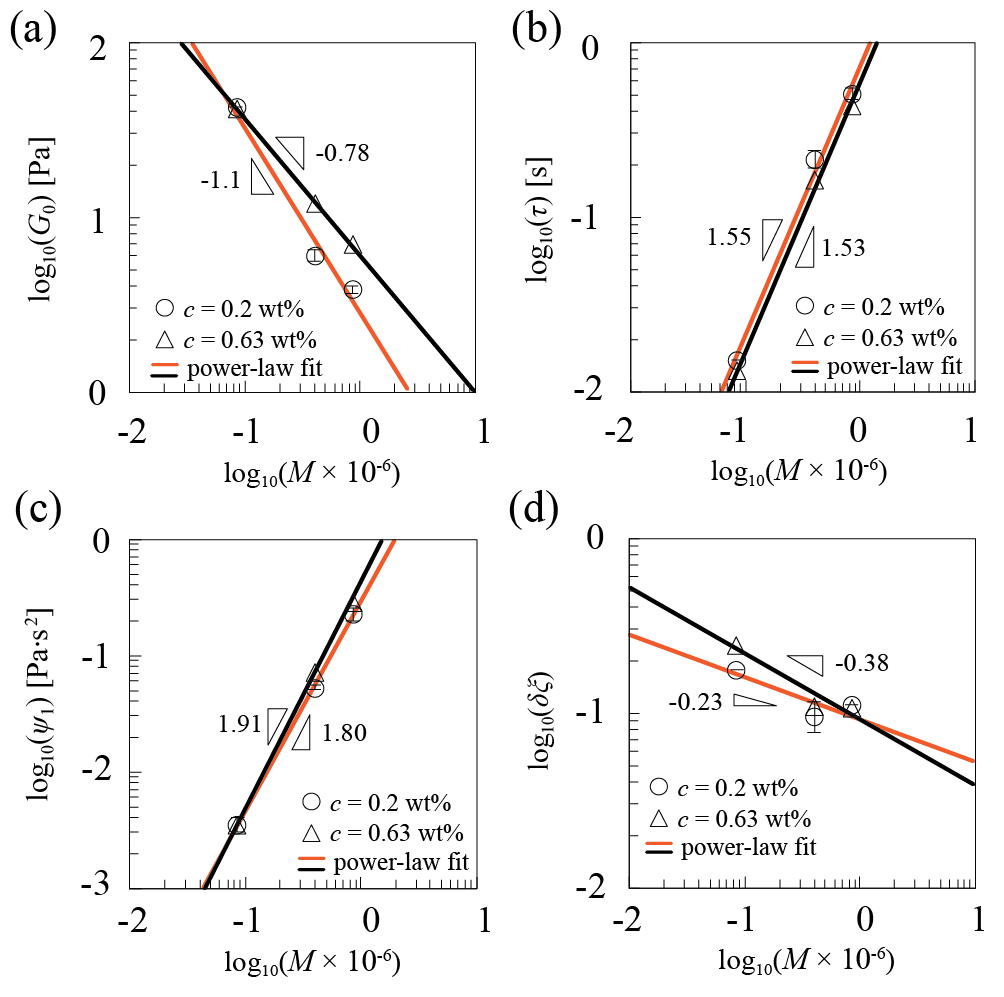}
\caption{\label{fig:M-dependence}Measured power-law dependence of a) $G_0$, b) $\tau$, c) $\psi_1$, and d) $\delta\xi$ on $M$, for constant $c=[0.2,0.63]$ wt\% and constant $\eta_s\approx 10.3$ Pa$\cdot$s. Refer to Tables~\ref{tab:summary table} and \ref{tab:summary table exp} for theory/experiment comparison.}
\end{figure}
Next, we investigate the dependence of $G_0, \tau$,  $\psi_1$, and $\delta\xi$ on the polymer molecular weight $M$. The experimentally measured properties are shown in Fig.~\ref{fig:M-dependence}. For $G_0$, the theoretical prediction is $G_0=k_BT\phi/(b^3N)\propto N^{-1}\propto M^{-1}$; i.e., $G_0$ is expected to decrease with increasing $M$. Intuitively, longer polymer chains are easier to bend (or deform). Experimentally, for two separate measurements, a slightly weaker dependence was observed, $G_0\propto M^{-0.93\pm0.15}$(Fig.~\ref{fig:M-dependence}(a)). 

Within the Zimm theory, the relaxation time is proportional to the cube of the radius of gyration $R_g$  of the polymer~\cite{gupta2000extensional,ewoldt2022designing}, $\tau\approx \eta_\text{s} R_g^3/(k_BT)$, where $R_g$ scales with the degree of polymerization $N$, such that $R_g\propto N^{\nu}\propto M^{\nu}$. Therefore, $\tau\propto M^{3\nu}$. In the asymptotic limits of long chains, $\nu=1/2$ in a $\theta$–solvent and $\nu\approx3/5$ in an athermal solvent. 

In \S\ref{sec:Zimm rheology}, we identified that the effective $\nu$ for our material is $\nu\approx0.47$. Therefore, we expect $\tau\propto M^{1.41}$. Our measurements closely confirm this trend, giving $\tau\propto M^{1.54\pm0.01}$~(Fig.~\ref{fig:M-dependence}(b)). Next, considering $\psi_1$, the theory predicts $\psi_1\sim G_0\tau^2\propto M^{6\nu-1}\approx M^{1.82}$, while the measurements give $\psi_1\propto M^{1.86\pm0.05}$~(Fig.~\ref{fig:M-dependence}(c)). We also find experimentally that $\delta\xi$ is $M$-dependent, such that $\delta\xi\propto M^{-0.31\pm0.07}$~(Fig.~\ref{fig:M-dependence}(d)).  In our design framework, we use the measured (fitted) exponents rather than the value of the exponents predicted by the Zimm theory.

It is important to note that the nominal values of $M$, provided by the manufacturer, do not reflect the measured values of the molecular weights (see Table~\ref{tab:molecular weight}). The discrepancies can depend on the proprietary production methods and the degradation under environmental stressors. To avoid these uncontrolled variances, we refer to the nominal $M$ of the products we obtained; while the exact power-law exponents may vary between different batches of the product, they can be readily identified from a few measurements, as we demonstrate in Fig.~\ref{fig:M-dependence}. In the Supplementary Material, we include how the result changes when $M_w$, not $M$, is used (see Table~\ref{tab:molecular weight} to find $M_w$).

\subsection{Dependence on solvent viscosity \label{sec:Solvent viscosity dependence}}
\begin{figure}
\includegraphics[width=85mm]{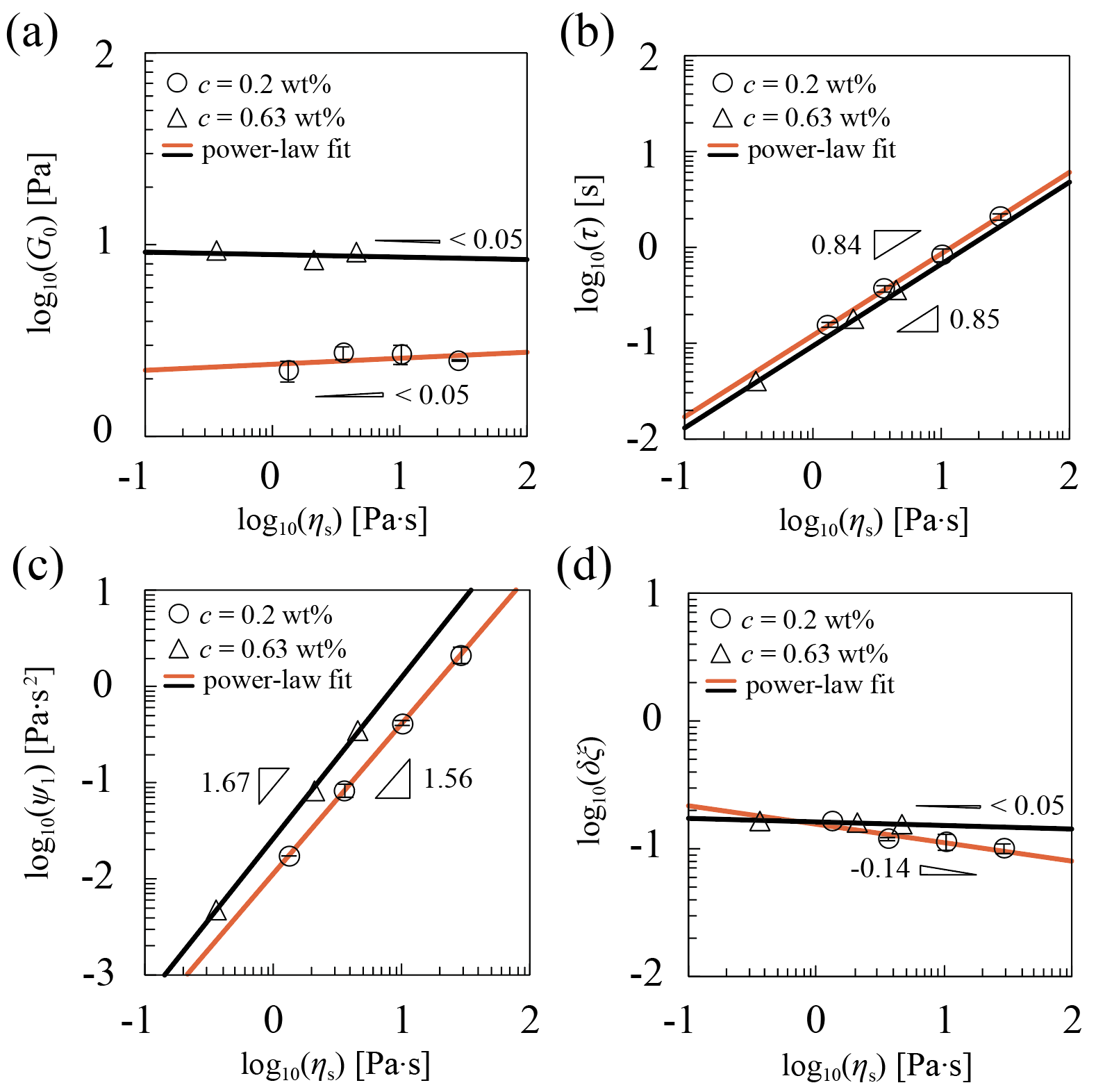}
\caption{\label{fig:eta-dependence} Measured power-law dependence of a) $G_0$, b) $\tau$, c) $\psi_1$, and d) $\delta\xi$ on $\eta_s$, for constant $c=[0.2,0.63]$ wt\% and constant $M= 1.27\times10^6$ . Refer to Tables~\ref{tab:summary table} and \ref{tab:summary table exp} for theory/experiment comparison.}
\end{figure}
Lastly, we varied the solvent viscosity $\eta_\text{s}$ by changing the weight fraction of PB in the PB/mineral oil mixture (see \S\ref{sec:Characterizing the dilute to semi-dilute transition}) to further test scaling relations presented in the preceding sections. First, $G_0=k_BT\phi/(Nb^3)$, which depends only on $c$ and $M$, and is therefore independent of $\eta_\text{s}$. This prediction is confirmed by the measurement shown in Fig.~\ref{fig:eta-dependence}(a). Next, since~\cite{zell2010there,bird1987dynamics,ewoldt2022designing} $\tau=\eta_s R_g^3/k_{B}T$, the relaxation time is expected to scale linearly with $\eta_s$. However, in our measurements we observed a weaker dependence, $\tau\propto \eta_s^{0.84\pm0.01}$, as shown in Fig.~\ref{fig:eta-dependence}(b). Next, since $\psi_1\propto G_0\tau^2$, the theory predicts $\psi_1\propto\eta_\text{s}^2$. Experimentally, we find $\psi_1\propto \eta_s^{1.61\pm0.05}$ (Fig.~\ref{fig:eta-dependence}(c)). Also, we found that $\psi_1/(\delta\xi)=\psi_{OB}\propto \eta_s^{1.69}$.

With these experimental measurements and the scaling arguments presented throughout \S\ref{sec:Concentration dependence} - \S\ref{sec:Solvent viscosity dependence}, we summarize the theoretical and experimental dependencies of  $G_0, \tau,  \xi(\nu)$, and $\psi_1$ on $c, M$, and $\eta_\text{s}$ in Tables~\ref{tab:summary table} and~\ref{tab:summary table exp}.

\begin{table}
\caption{\label{tab:summary table} Theoretical dependence of $G_0, \tau$, and $\psi_1$ on $c, M$, and $\eta_\text{s}$ (where $\nu\approx0.47$). These results can be derived from the scaling arguments presented in \S\ref{sec:Concentration dependence}- \ref{sec:Solvent viscosity dependence}.}
\begin{ruledtabular}
\begin{tabular}{ccccc}
 &$G_0$&$\tau$&$\psi_1$ &$\delta\xi$\\
\hline
$c$&  $\propto c^1$&  $\propto c^0$&$\propto c^1$ &N/A\\
$M$&  $\propto M^{-1}$& $\propto M^{3\nu\approx1.41}$&$\propto M^{6\nu-1\approx 1.82}$ &N/A\\
$\eta_\text{s}$&  $\propto \eta_\text{s}^{0}$& $\propto \eta_\text{s}^{1}$&$\propto \eta_\text{s}^{2}$ &N/A\\
\end{tabular}
\end{ruledtabular}
\end{table}
\begin{table}
\caption{\label{tab:summary table exp} Observed (measured) power-law dependence of $G_0, \tau$,  $\psi_1$, and $\delta\xi$ on $c, M$, and $\eta_\text{s}$, with two significant digits after the decimal. The values are based on the two sets of data that are shown in Figs.~\ref{fig:c-dependence}, \ref{fig:M-dependence}, and \ref{fig:eta-dependence}. }
\begin{ruledtabular}
\begin{tabular}{ccccc}
 &$G_0$&$\tau$&$\psi_1$ &$\delta\xi$\\
\hline
$c$&  $\propto c^{1.00\pm0.003} $&  $\propto c^{0.10\pm0.01}$&$\propto c^{1.41\pm0.06}$&$\propto c^{0.20\pm0.07}$\\
$M$&  $\propto M^{-0.93\pm0.15}$& $\propto M^{1.54\pm0.01}$&$\propto M^{1.86\pm0.05}$&$\propto M^{-0.31\pm0.07}$\\
$\eta_\text{s}$&  $\propto\eta_s^{0.01\pm0.02}$& $\propto \eta_\text{s}^{0.84\pm0.01}$&$\propto \eta_\text{s}^{1.61\pm0.05}$&$\propto \eta_\text{s}^{-0.08\pm0.06}$\\
\end{tabular}
\end{ruledtabular}
\end{table}
\section{Design fluids with controlled properties}\label{sec:Designing PIB Boger fluids with desired mechanical properties}
\subsection{Design matrix}
Note that the power-law relationships presented in Table~\ref{tab:summary table} can be re-written as (with $\nu=0.47$) 
        \begin{equation} \label{eqn: matrix form}
        \renewcommand{\arraystretch}{1.3}
        \left(
        \begin{array}{ccc}
                1  &  -1        &  0               \\
                0  &  1.41   & 1     \\
                1 &  1.82 &  2                \\
                
        \end{array}  \right)
        \left(\begin{array}{c}
                \log_{10} c      \\
                \log_{10} M      \\
                \log_{10} \eta_\text{s}      \\
        \end{array} \right) \propto 
        \left(\begin{array}{c}
                \log_{10} G_0      \\
                \log_{10} \tau      \\
                \log_{10} \psi_1      \\
        \end{array} \right).
        \end{equation}
The determinant of the coefficient matrix on the left-hand side of Eq.~(\ref{eqn: matrix form}) is zero, so the columns are linearly dependent. This outcome is expected, as $\psi_1$ is determined by $G_0$ and $\tau$. This implies that one property can remain constant while the other two are varied simultaneously.

Instead, the experimental measurements shown in Figs.~\ref{fig:c-dependence}, \ref{fig:M-dependence}, and \ref{fig:eta-dependence} and Table~\ref{tab:summary table exp} indicate otherwise. Converting the numbers in Table~\ref{tab:summary table exp} into matrix form yields the following equation (rounded to two decimal places, and ignoring the ranges),
        \begin{equation} \label{eqn: matrix form exp}
        \renewcommand{\arraystretch}{1.3}
        \textbf{C}\approx\left(
        \begin{array}{ccc}
                1.00     &  -0.93  &  0.01               \\
                0.10  &  1.54  & 0.84     \\
                1.41 &  1.86    &  1.61                \\
        \end{array}  \right),~
        \textbf{C}\cdot\left(\begin{array}{c}
                \log_{10} c      \\
                \log_{10} M      \\
                \log_{10} \eta_\text{s}      \\
        \end{array} \right) \propto 
        \left(\begin{array}{c}
                \log_{10} G_0      \\
                \log_{10} \tau      \\
                \log_{10} \psi_1      \\
        \end{array} \right).
        \end{equation}

Each row of this equation contains quantities of mixed dimensions, as indicated by the use of the proportionality symbol `$\propto$’. To simplify notation and ensure dimensional consistency, we use overbar (`$\overline{\cdot}$') symbol to denote ratios between properties in a new fluid (to be designed) relative to those of a reference fluid, e.g., $\overline{M}=M_{\text{new}}/M_{\text{ref}}$. This new notation makes it easier to understand how properties in a new fluid differ from those of a fluid of reference. Then, we compute 
\begin{equation}\label{eqn:deriving dimensionless equation}
\begin{aligned}
(\log_{10} \overline{G},& ~\log_{10} \overline{\tau},\log_{10} \overline{\psi})^{\text{T}}\\
=&(\log_{10} G_\text{0,new},\log_{10} \tau_\text{new},\log_{10} \psi_\text{1,new})^{\text{T}}\\
&-(\log_{10} G_\text{0,ref},\log_{10} \tau_\text{ref},\log_{10} \psi_\text{1,ref})^{\text{T}} 
\end{aligned}
\end{equation}
and invert the expression to write 
\begin{align} \label{eqn: matrix form exp inverted}
        \renewcommand{\arraystretch}{1.3} 
        \textbf{C}^{-1}&\approx\left(
        \begin{array}{ccc}
                -19.277     &  -31.590  &  16.624               \\
                -21.490  &  -33.467  & 17.618    \\
                41.508 &  66.009   &  -34.123                \\
        \end{array}  \right),\nonumber \\~
        \left(\begin{array}{c}
                \log_{10} \overline{c}      \\
                \log_{10} \overline{M}      \\
                \log_{10} \overline{\eta}_s      \\
        \end{array} \right)& = 
        \textbf{C}^{-1}\cdot
        \left(\begin{array}{c}
                \log_{10} \overline{G}      \\
                \log_{10} \overline{\tau}      \\
                \log_{10} \overline{\psi}      \\
        \end{array} \right).
\end{align}
Since the design matrix $\textbf{C}$ has full rank, and is therefore invertible, there is a unique mapping between $\overline{G}, \overline{\tau},$ and $\overline{\psi}$ to $\overline{c}, \overline{M},$ and $\overline{\eta}_s$, and vice versa. As an example, if one wants to design fluids, with constant $G_0$ and $\tau$ (i.e., $\overline{G}=\overline{\tau}=1$) but variable $\psi_1$ (e.g., $\overline{\psi}=0.1$), one can compute $\textbf{C}^{-1}\cdot(0,0,-1)^\text{T}$ to determine the corresponding values of $\overline{c}, \overline{M},$ and $\overline{\eta}_s$, thereby yielding the target fluid recipe.  Given Eq.~(\ref{eqn:deriving dimensionless equation}), the matrix equation Eq.~(\ref{eqn: matrix form exp inverted})  is dimensionless, and we stress that the these dimensionless numbers inside $\textbf{C}^{-1}$ are only valid for the PIB-PB Boger fluid system we studied. For different model Boger fluid systems, these numbers will be different, though can be measured with the process shown in earlier sections.

\subsection{$G$, $\tau$, and/or $\psi_1$-controlled fluids}
Using Eq.~(\ref{eqn: matrix form exp inverted}), we now can design fluids with desired viscoelastic properties. To demonstrate, we fabricated five fluid, whose target $\overline{G}$ and $\overline{\tau}$ are listed in Table~\ref{tab:test fluids design}, with labels A through E. We chose an arbitrary set of values for $c$, $M$, and $\eta_s$ for the fluid A, which we use as reference. With Eq.~(\ref{eqn: matrix form exp inverted}), we then derived the fluid recipe for the desired target properties. Specifically, we reduced $G_0$ of fluid B and C relative to that of the fluid A, such that $G_{0,B}=0.5\times G_{0,A}$ and $G_{0,C}=0.1\times G_{0,A}$. Similarly, we varied the values of $\tau$ of fluid D and E relative to that of the fluid A, such that $\tau_{B}=0.5\times \tau_{A}$ and $\tau_{C}=0.1\times \tau_{A}$.

\begin{figure*}
\includegraphics[width=170mm]{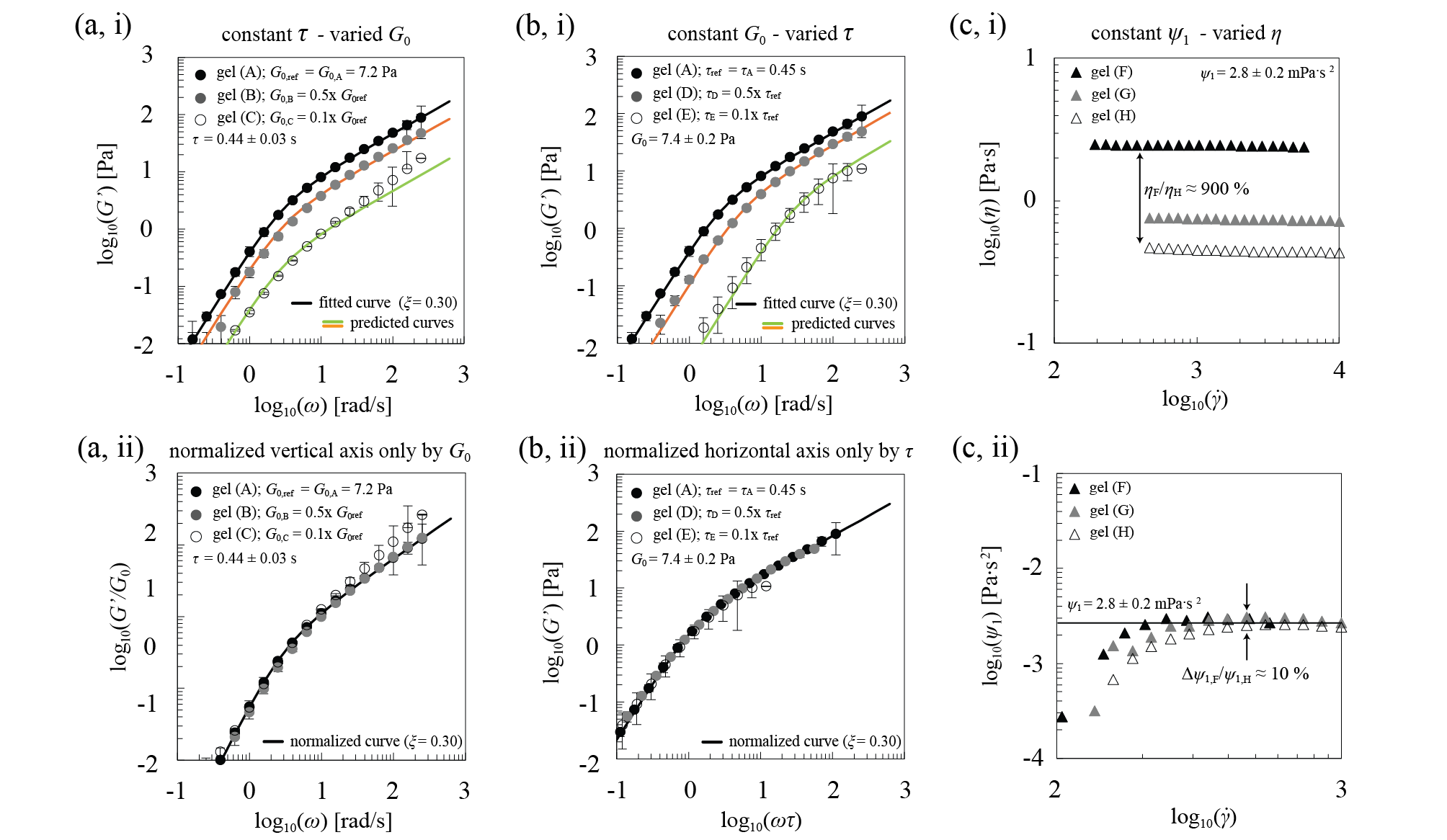}
\caption{\label{fig:G-tau-varied design fluids} Experimental verification of the proposed design scheme in Eq.~\ref{eqn: matrix form exp inverted}. 
a) The fluids A, B, and C are designed to share an identical $\tau$ but has varied $G_0$, such that $G_{0,B}=0.5\times G_{0,A}$ and $G_{0,C}=0.1\times G_{0,A}$. Collapse of $G'$ data in panel (\textit{ii}) normalized by the measured $G_0$ indicate that the design was successful with marginal error. 
b) Likewise, the fluids A, D, and E should instead share a constant $G_0$ but have varied $\tau$ such that $\tau_{B}=0.5\times \tau_{A}$ and $\tau_{C}=0.1\times \tau_{A}$. In panel (\textit{ii}), we show collapsed curves of $G'$ plotted against normalized horizontal axis, $\omega\tau$, indicating the three fluid share nearly the same $G_0$ but have distinct values of $\tau$. 
c) Lastly, fluids F, G, and H are designed to share the same $\psi_1$. To highlight that the viscosity and elasticity can be successfully decoupled, we show in (\textit{i}) $\eta$ of fluids F, G, and H, which ranges nearly a decade, and in (\textit{ii}) $\psi_1$ of fluids F, G, and H, whose variance is considerably small compared to that of $\eta$. 
For the specifics of the fluids A through E, refer to Tables~\ref{tab:test fluids design} and \ref{tab:test fluids fabricated}. And for fluids F, G, and H, refer to Tables ~\ref{tab:test fluids design psi} and \ref{tab:test fluids fabricated psi}. }
\end{figure*}

We then measured the values of $G_0$ and $\tau$ for fluids A through E, as reported in Fig.~\ref{fig:G-tau-varied design fluids}(a, b) and tabulated in the corresponding columns in Table~\ref{tab:test fluids fabricated}. 
\begin{table}
\caption{\label{tab:test fluids design} Design of Boger fluids with systematically varied $G_0$ and $\tau$. With the fluid compositions calculated with Eq.~(\ref{eqn: matrix form exp inverted}), we fabricated fluid A through E to vary $G_0$ between the fluid A, B, and C, and to vary $\tau$ between the fluid A, D, and E. The composition for the fluid A is arbitrarily chosen to provide a reference point. }
\begin{ruledtabular}
\begin{tabular}{cccccc}
 & \multicolumn{2}{c}{Target}& \multicolumn{3}{c}{Design (Eq.~\ref{eqn: matrix form exp inverted})}\\\toprule
 &  $\overline{G}$& $\overline{\tau}$& $c$ [wt\%]&$M\times10^{-6}$& $\eta_\text{s}$ [Pa$\cdot$s]\\\midrule
 \hline
 A (ref)&  1& 1& 0.63&1.27& 4.57\\
 B&  0.5& 1& 0.32&1.27& 4.96\\
 C&  0.1& 1& 0.063&1.27& 5.97\\
 D&  1& 0.5& 0.64&1.27& 2.01\\
 E&  1& 0.1& 0.65&1.27& 0.30\\ \bottomrule \end{tabular}
\end{ruledtabular}
\end{table}
\begin{table*}
\caption{\label{tab:test fluids fabricated}Measured/fitted rheological properties in fluids A through E. Targeted (see Table~\ref{tab:test fluids design}) and measured $\overline{G}$ and $\overline{\tau}$ are within marginal differences between each other, demonstrating efficacy of our design scheme in Eq.~(\ref{eqn: matrix form exp inverted}). See Fig.~\ref{fig:G-tau-varied design fluids}(a,b) also. }
\begin{ruledtabular}
\begin{tabular}{cccccccc}
 \midrule
 & $G_0$ [Pa] &$\tau$ [s]  &   $\eta_\text{s}$ [Pa$\cdot$s]&$\psi_1$ [Pa$\cdot$s$^{2}$]&$\overline{G}$&$\overline{\tau}$&$\overline{\psi}$\\\midrule
\hline
A (ref)& 7.26&0.45&   4.57&0.34&1&1&1\\
B& 3.27&0.43&   4.89&0.14&0.45&0.94&0.36\\
C&  0.676& 0.44&   5.85&0.018&0.09&0.98&0.05\\
 D& 7.23& 0.24&   2.10&0.083&1.00&0.52&0.21\\
 E& 7.22& 0.053&  0.33&0.0048& 0.99& 0.12&0.01\\ \bottomrule \end{tabular}
\end{ruledtabular}
\end{table*}
Consistent with our predictions, $G_0$ of the fluid B and C were respectively $0.45\times$ and $0.09\times$ to that of the reference fluid A while $\tau$ of the fluid D and E were respectively $0.52\times$ and $0.12\times$ that of A. These measurements are shown in Fig.~\ref{fig:G-tau-varied design fluids}(a,b), respectively. In particular, when we normalized $G'$ of fluid A, B, and C by $G_0$, along the vertical direction without modifying the horizontal axis, the three lines in the top panel of Fig.~\ref{fig:G-tau-varied design fluids} collapse onto a single line (see lower panel of Fig.~\ref{fig:G-tau-varied design fluids}(a)), confirming that $\tau$ is held relatively constant while $G_0$ varied between the fluids. Likewise, when we normalized $\omega$ with $\tau$ of fluid A, D, and E, without altering the vertical axis, the three lines shown in the top panel of  Fig.~\ref{fig:G-tau-varied design fluids}(b) collapsed onto a single line (see lower panel). This result shows that $G_0$ between the three fluids remained relatively constant. These measurements testify to the success of our approach in controlling $G_0$ and $\tau$ independently. 

Lastly, we fabricated three additional fluids, labeled F, G, and H where F is the reference fluid, to test if $\psi_1$ can be controlled similarly. Table~\ref{tab:test fluids design psi} lists the compositions used for the fluid F, G, and H. These formulations will create fluid with varied $G_0$ and $\tau$, but $\psi_1$ is predicted to remain constant relative to that of the reference fluid F. Since $\psi_1$ often reflects the effect of polymer elasticity in the fluid, we highlight that the solvent viscosities $\eta_s$ between the test fluids were varied substantially across the test fluids. In practice, fluids like F, G, and H can be used to clearly decouple the effect of fluid elasticity from those of fluid viscosity. 

Measured values of $\psi_1$ for fluids F, G, and H are given in Table~\ref{tab:test fluids fabricated psi} and in Fig.~\ref{fig:G-tau-varied design fluids}(c). $\psi_{1,G}$ was  $\sim5\%$ higher and $\psi_{1,H}$ was $\sim10\%$ lower than the target values. These fluids are useful as the fluid viscosities varied by a factor of $\approx 9$ between the lowest and the highest. Likewise, fluids with constant $\eta_s$ and varied $\psi_1$ can be fabricated  using Eq.~(\ref{eqn: matrix form exp inverted}) as a guide.
\begin{table}
\caption{\label{tab:test fluids design psi} Design of Boger fluids with systematically controlled $\psi_1$. With the fluid compositions calculated with Eq.~(\ref{eqn: matrix form exp inverted}), we fabricated fluid F, G, and H to hold $\psi_1$ between the fluids constant while varying $\eta$. Here, fluid F is used as a reference.}
\begin{ruledtabular}
\begin{tabular}{ccccc}
 & Target& \multicolumn{3}{c}{Design (Eq.~\ref{eqn: matrix form exp inverted})}\\\toprule
 &  $\overline{\psi}$& $c$ [wt\%]&$M\times10^{-6}$& $\eta_\text{s}$ [Pa$\cdot$s]\\\midrule
 \hline
 F (ref)&  1& 0.15&0.4& 2.37\\
 G&  1& 0.29&0.85& 0.56\\
 H&  1& 0.42&1.27& 0.26\\ \end{tabular}
\end{ruledtabular}
\end{table}
\begin{table}
\caption{\label{tab:test fluids fabricated psi} Measured $\overline{\psi}=\psi_1/\psi_{1,\text{ref}}$ of fluids F, G, and H. We performed rotational measurements only for these fluid (hence no $G_0$ and $\tau$). The viscosities between the fluids vary nearly up to 900\% between the reference and the smallest, while variances in $\overline{\psi}$ are nearly negligible in comparison. This set of fluids can be used in experiments that aim to decouple viscous effects from elastic effects.}
\begin{ruledtabular}
\begin{tabular}{cccc}\toprule
 &$\eta$ [Pa$\cdot$s]& $\psi_1$ [mPa$\cdot$s$^{2}$]&$\overline{\psi}$\\\midrule
 \hline
 F (ref)&2.49& 2.85&1\\
 G&0.76& 2.99&1.05\\
 H&0.47& 2.54&0.89\\ \end{tabular}
\end{ruledtabular}
\end{table}

\section{Conclusion}
In this work, we integrate known concepts from polymer physics and rheology to provide a means for creating well-controlled dilute polymer solutions with prescribed rheological properties. The framework we developed for a PIB-PB based dilute viscoelastic fluid is based on experimental measurements of rheological properties using a conventional rheometer. With a few initial measurements (e.g., identifying the linearity limit where $c=c^*$) and identifying the power-law relationship between the input parameters and the output properties, our framework can further be extended to design the viscoelastic properties of various Boger fluids, whether they are made with polyisobutylene, polyacrylamide, polystyrene, or some other types of flexible Gaussian polymer chains.

The methodology demonstrated here for designing viscoelastic fluids provides intuition for the observations in various viscoelastic flow experiments. For example, one can design different fluids with either constant $\eta_s$ and varied $\psi_1$ or constant $\psi_1$ and varied $\eta_s$ to examine if the source of the observed viscoelastic behavior originates from the effects associated with the fluid viscosity or the fluid elasticity. Furthermore, this method can be used to estimate the change in molecular weight of the polymers after degradation or scission by using $\overline{G}, \overline{\tau},$ and $\overline{\psi}$ as inputs and $\overline{c}, \overline{M},$ and $\overline{\eta}_s$ as outputs using Eq.~(\ref{eqn: matrix form exp inverted}). This capability is made possible by the non-ideality of the actual fluids, which allows a  linear mapping between $\overline{G}$, $\overline{\tau}$, $\overline{\psi}$ and $\overline{c}$, $\overline{M}$, $\overline{\eta}_s$. 

Lastly, we suggest a couple of avenues for future work to improve the methodology introduced here. First, a design scheme that depends less on the quality of fitting will make our idea more robust. Our method relies on the fitting to identify the power-law exponents in the matrix equation. However, torque limitations in a conventional shear rheometer, combined with the polydispersity of commercially sourced polymers, can reduce the reliability of the fitting. We found that applying the time-temperature superposition principle to avoid the low-torque limit is generally not helpful here, because changing temperature does not translate the curve in the vertical direction (i.e., to have higher $G'$ values) and also affects $\xi$. Second, $\psi_1$ is not constant over the full range of $\dot\gamma$ but rather decreases at high $\dot\gamma$. Such decrease in $\psi_1$ is hard to predict and requires numerous fitting parameters~\cite{das2024laun}. However, given recent advances in the machine learning techniques and their applicability in understanding viscoelastic flows~\cite{lennon2023scientific}, we believe gaining control of $\psi_1$ over the wider range of $\dot\gamma$ is possible if machine learning is combined with our design scheme.

In conclusion, we presented a practical framework for designing dilute viscoelastic fluids with specified $G_0$, $\tau$, and $\psi_1$ using readily available materials. This approach enables tailored Boger fluids for flow experiments while providing insight into how concentration, molecular weight, and solvent viscosity shape viscoelastic behavior. 

\section*{SUPPLEMENTARY MATERIAL}
See the supplementary material for more details of the GPC measurement, additional discussions around the fitting scheme, and a note on the long-term thermal stability of the sample stocks.

\begin{acknowledgments}
We thank Nan Hu, J. Pedro de Souza, Tachin Ruangkriengsin, and Charles Schroeder for helpful discussions and Cristina Preston-Herrera and Erin Stache for help with GPC measurements. For financial support, we acknowledge US National Science Foundation (NSF), specifically grant no. CBET-2246791 and the Princeton Materials Research Science and Engineering Center (MRSEC, DMR-2011750). J. H. thanks Kwanjeong Educational Foundation Graduate Fellowship for funding.\\
\end{acknowledgments}

\section*{AUTHOR DECLARATIONS}
\subsection*{Conflicts of interest}
There are no conflicts to declare.
\subsection*{Data availability}
The data that support the findings of this study are available within the article and its supplementary material.

\section*{REFERENCES}
%

%
\clearpage
\onecolumngrid

\appendix
\renewcommand{\thesection}{\Roman{section}}

\renewcommand{\appendixname}{Supplementary Note}
\renewcommand{\thefigure}{S\arabic{figure}}
\setcounter{figure}{0}

\begin{center}
{\LARGE Supplementary Material for ``Design of model Boger fluids with systematically controlled viscoelastic properties''}\\[2em]
{Jonghyun Hwang and Howard A. Stone}\\
{Department of Mechanical and Aerospace Engineering, Princeton University, Princeton, NJ, 08544, USA}
\end{center}

\vspace{2em}

\section{Details on the gel permeation chromatography (GPC) measurement}
\begin{figure*}
\includegraphics[width=160mm]{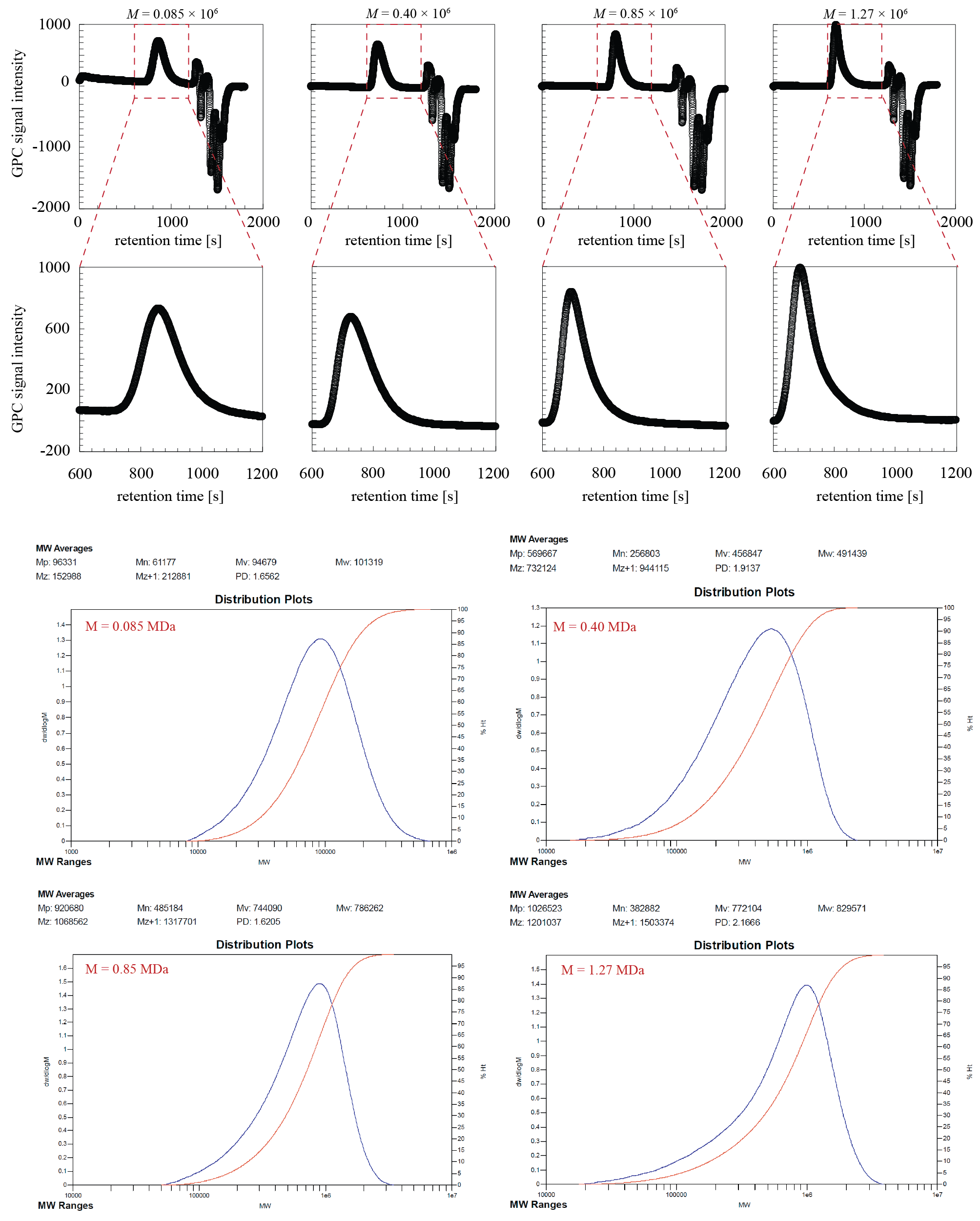}
\caption{\label{fig:GPC} Molecular weight measurement using gel permeation chromatography (GPC). Top 8 figures correspond to the raw data of signal intensity versus retention (run) time extracted from the machine (Infinity, Agilent), and the bottom 4 figures are snapshots of the molecular weight distribution that were directly imported from the report generated by the machine. $M$ corresponds to the nominal value reported by Scientific Polymer Products Inc. }
\end{figure*}
Using the gel permeation chromatography technique, we identified the actual values of the molecular weights of the polyisobutylene (PIB) products that were sourced from Scientific Polymer Products Inc. We used the Infinity series machine from Agilent, with the following calibration curve: $\log(M_w)=15.672444-1.450312 x^1+0.068374 x^2-0.001344 x^3$, where $x$ is the retention time and the calibration curve is defined between the two limits of molecular weights: $1,000<M_w<1,000,000 $ MDa. The measurements were performed at 35 $^\circ$C. The PIB samples of different nominal molecular weights were dissolved in tetrahydrofuran at 1 mg/ml and were filtered with 0.2 $\mu$m PTFE filter. The raw GPC data is shown in Fig.~\ref{fig:GPC}.

We note that the nominal value of the $M=1.27\times10^6$  sample falls outside the calibration range ($M=10^3$ to $10^6$). Despite the measured $M_w\approx 0.830\times10^6$ for that sample, a portion of the chains longer than $M=1\times10^6$ may not have been correctly detected. This difference may be related to the larger than expected variability displayed in Fig. 7(c).

\section{Nominal $M$ versus measured $M_w$ when identifying the power-law exponents }
\begin{figure*}
\centering
\includegraphics[width=160mm]{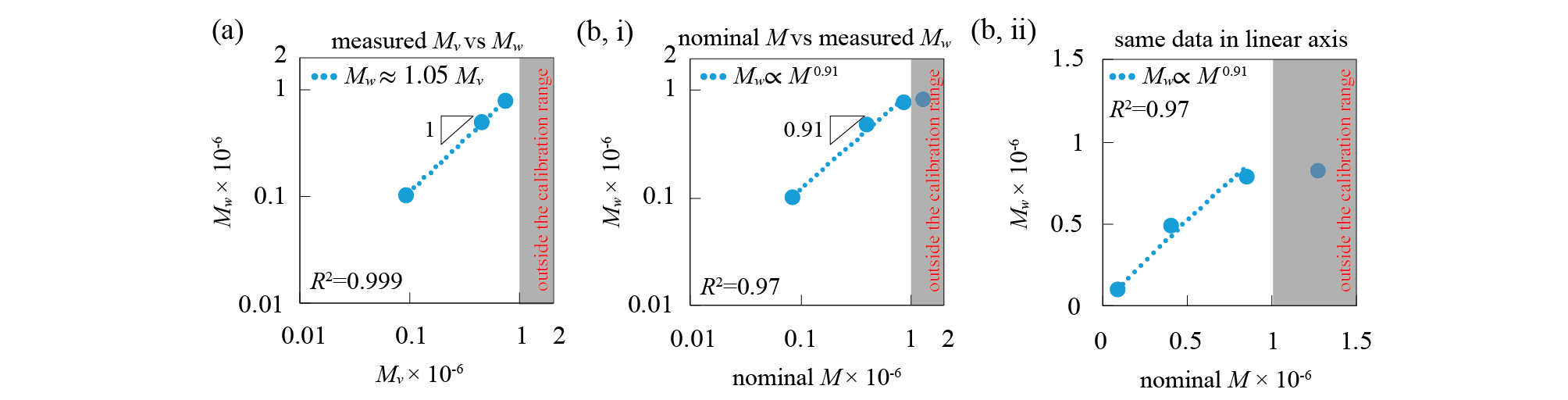}
\caption{\label{fig:M vs Mw} Relationship between the nominal PIB viscosity-averaged molecular weight (M) and the measured weight-averaged molecular weight. We find that the measured viscosity-averaged molecular weight $M_v$ and $M_w$ are nearly identical. On the other hand, the relationship between $M_w$ and $M$ can be approximated as $M_w\propto M^{0.91}$. }
\end{figure*}
We discussed in the main article that the nominal values of the molecular weights ($M$) of the PIB products differ from the values obtained from the GPC. Here, we show that using the measured $M_w$ instead of $M$ likewise results in a viable design matrix equation Eq. (10). The result show that the numbers on the center column of the coefficient matrix $\textbf{C}$ is adjusted with a factor of $1/0.91$.

Noticing that the nominal $M$ values are the viscosity-averaged molecular weights ($M_v$), as stated by the manufacturer, we first identify how the measured $M_v$ and $M_w$ compare with each other (Fig.~\ref{fig:M vs Mw}(a)). We found that our PIB samples show that $M_w\approx 1.05 M_v$. Therefore, we conclude that when we compute $\log_{10}\overline{M}$ as we do in the main text, $M_w$ and $M_v$ can be interchangeably used. 

Next, we analyze how the nominal $M$ is compared with $M_w$, as shown in Fig.~\ref{fig:M vs Mw}(b). By fitting, we obtained that $M_v\propto M^{0.91}$, and likewise $M_w\propto M^{0.91}$. Therefore, $\log_{10}M=\frac{1}{0.91}\log_{10}M_w$. With this, Eq. (10) in the main text can be re-written as
\begin{equation} \label{eqn: matrix form exp}
        \renewcommand{\arraystretch}{1.3}
        \textbf{C}\approx\left(
        \begin{array}{ccc}
                1.00     &  -1.02  &  0.01               \\
                0.10  &  1.69  & 0.84     \\
                1.41 &  2.04    &  1.61                \\
        \end{array}  \right),~
        \textbf{C}\cdot\left(\begin{array}{c}
                \log_{10} c      \\
                \log_{10} M      \\
                \log_{10} \eta_\text{s}      \\
        \end{array} \right) \propto 
        \left(\begin{array}{c}
                \log_{10} G_0      \\
                \log_{10} \tau      \\
                \log_{10} \psi_1      \\
        \end{array} \right).
        \end{equation}

\section{Comparing the fitted $G_0$ to the dumbbell estimate $G_0\sim cRT/M_w$}
\begin{figure}
    \centering
    \includegraphics[width=0.5\linewidth]{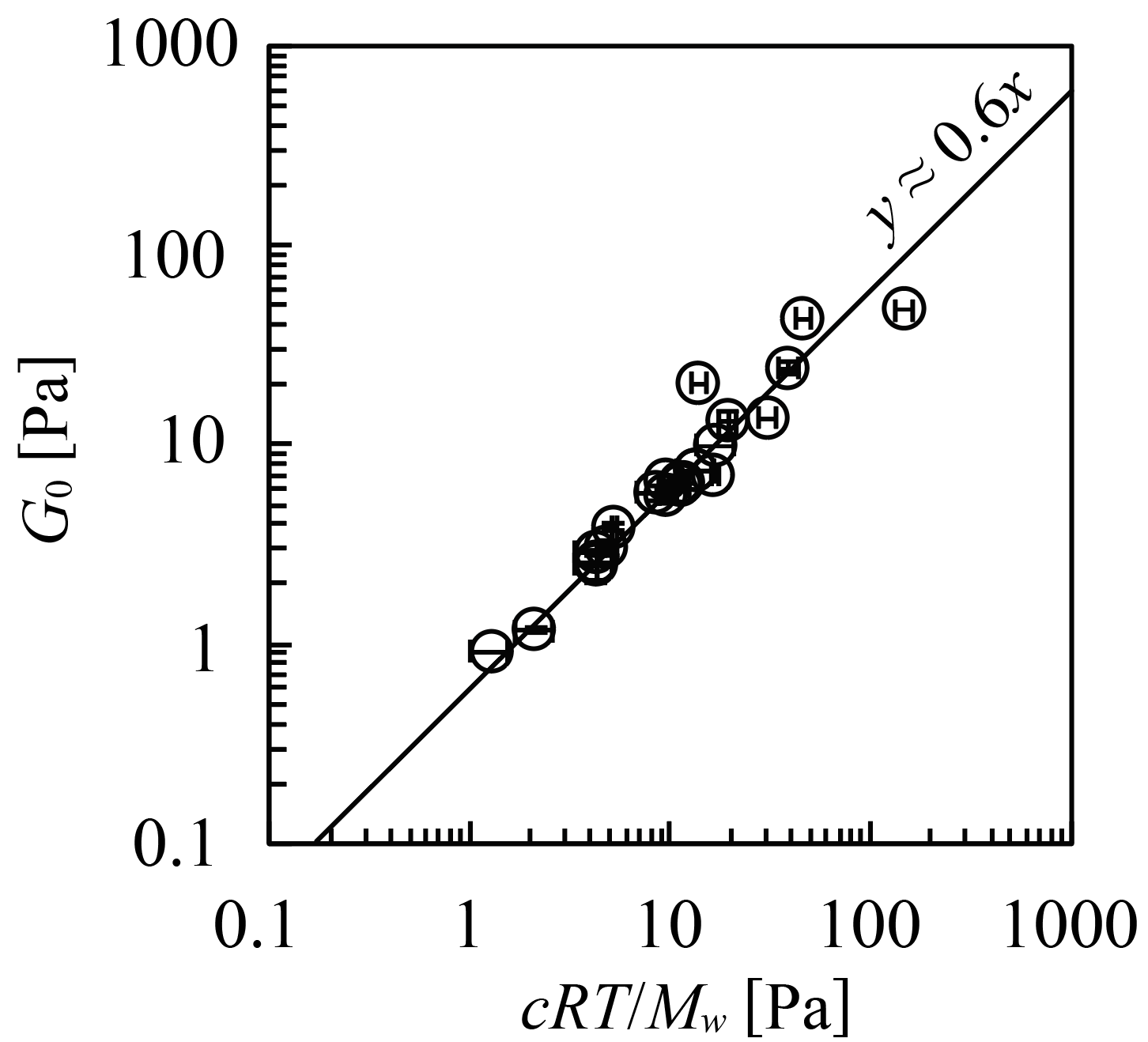}
    \caption{Fitted $G_0$ compared with the dumbbell model definition of $G_0\equiv cRT/M_w$, where $[c]=\text{kg/m}^3$, $[RT]=$J/mol, and $[M_w]=$ kg/mol.}
    \label{fig:G0permol}
\end{figure}
The shear modulus of a polymeric fluids are often related to the `thermal energy ($k_BT$) per number of chains'-type of argument. Specifically, models that represent a polymer as an entropic spring that obey Hookean response defines the shear modulus as $G_0=cRT/M_w$. In Fig.~\ref{fig:G0permol}, we compare the fitted $G_0$ and $cRT/M_w$, which shows that the two quantities are linearly proportional to each other over the two decades. In the calculation, we treat $M_w$ a value expressed in kg/mol quantity.

\section{Additional characterization for the $G_0-$ and $\tau-$controlled design fluids}
\begin{figure}
    \centering
    \includegraphics[width=0.5\linewidth]{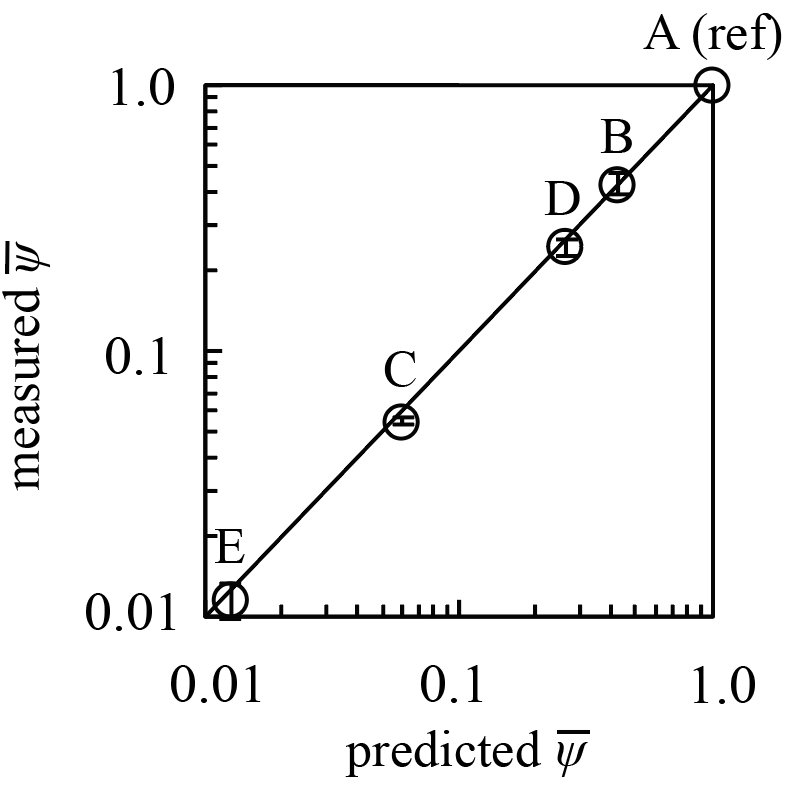}
    \caption{A supplement to the contents in Table V in the main text. For the fluids A to E, this plot compares the predicted and the measured values of $\overline{\psi}$.}
    \label{fig:Fig_FluidsAtoE_psi_comparison}
\end{figure}
When fabricating the fluids A to E (also see Table V in the main text), we fixed the values for $\overline{G}, \overline{\tau}$, and $\overline{M}$. Specifically, $\overline{M}$ was held to equal 1 for all of the fluids A to E, and $\overline{G}$ and $\overline{\tau}$ took values of $1, 0.5$, or $0.1$. The design equation linearly maps three variables to another three variables. Having three variables ($\overline{G}, \overline{\tau}$, and $\overline{M}$) fixed, one can identify the values of the remaining three variables ($\overline{c}, \overline{\eta}$ and $\overline{\psi}$) that satisfy the design equation (Eq. 12). In other words, we can predict the change in $\psi$ of the fluids B to E relative to the fluid A. In Fig.~\ref{fig:Fig_FluidsAtoE_psi_comparison}, we compared the prediction with the measured $\overline{\psi}$, which demonstrates that the usefulness of the design equation.

\section{Thermal degradation of PIB in the long-term mixing at an elevated temperature }
\begin{figure}
    \centering
    \includegraphics[width=1\linewidth]{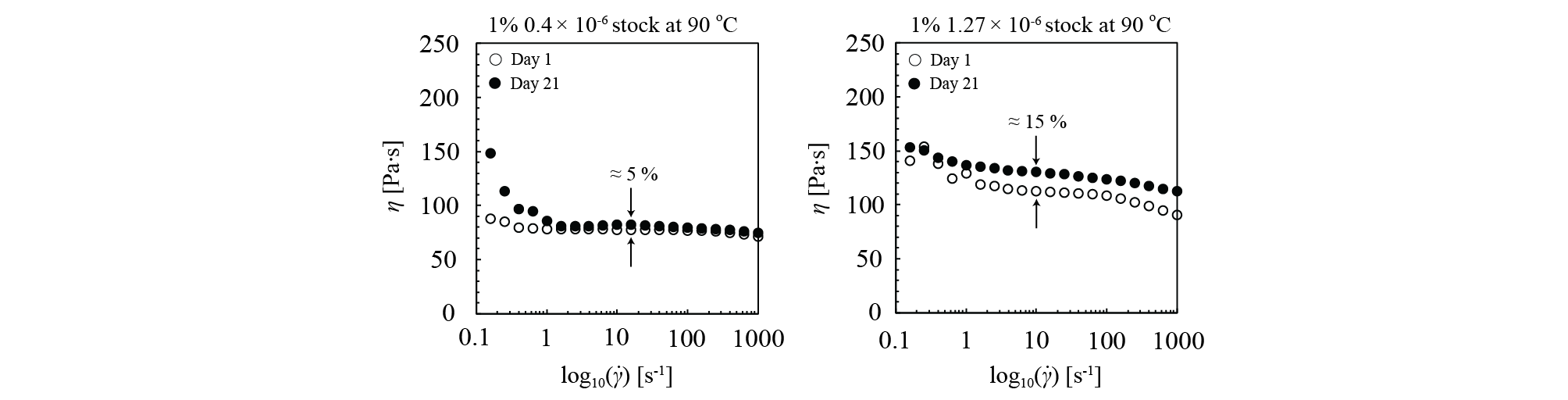}
    \caption{Long-term storage and the thermal stability of the 1\% PIB in light mineral oil. We assessed thermal degradation of the samples by measuring their viscosities. $M=$ 0.4 and 1.27 $\times10^6$ stocks, stirred for 3 weeks at 90~$^\circ$C, showed viscosities reduced by approximately 5\% and 15\%, respectively. Two repeats were averaged and the error bars are negligibly small.}
    \label{fig:Fig_long-term-storage}
\end{figure}
The sample preparation method involves dissolution of PIB in mineral oil at an elevated temperature (80 $^\circ$C) for a prolonged period of time. To estimate the thermal degradation of PIB during this process, we measured viscosities of 1\% stock solutions of 0.4 and 1.27 $\times10^6$ PIB in light mineral oil (Fig.~\ref{fig:Fig_long-term-storage}). The stock solutions were kept at $90~^\circ$C for three weeks, constantly stirring on a magnetic stirrer. We observed that, over the three-week period, solution viscosities drop by $\sim5\%$ for $M=0.4\times10^6$  and $\sim15\%$ for $M=1.27\times10^6$  stocks, suggesting that longer chains are more unstable. As we prepared our stocks at a lower temperature for 3 - 10 days, depending on the length of the chains, we believe the thermal degradation is sufficiently low.

\end{document}